\documentclass[12pt,a4paper]{article}

\usepackage[left=1in, right=1in, top=1in, bottom=1in, twoside=false]{geometry}
\usepackage[onehalfspacing]{setspace}
\usepackage[]{natbib}

\usepackage[dvips]{graphicx}
\usepackage[dvipsnames]{xcolor}

\usepackage[colorlinks=true,linkcolor=blue, citecolor=blue]{hyperref}

\usepackage{amsmath, amsthm, amssymb,amsfonts}
\usepackage{mathtools}
\usepackage{soul}
\usepackage{tikz}
\usepackage{comment}

\newtheorem{prop}{Proposition}

\newtheorem{lmm}{Lemma}

\theoremstyle{definition}
\newtheorem{example}{Example}

\newtheorem{defi}{Definition}

\usepackage{etoolbox}
\AtEndEnvironment{example}{\null\hfill\qed}

\title{Time-varying Cost of Distancing: \\ Distancing Fatigue and Lockdowns}
\author{Christoph Carnehl\thanks{Bocconi University, Department of Economics and IGIER. Email: \texttt{christoph.carnehl@unibocconi.it}.} \and
Satoshi Fukuda\thanks{Bocconi University, Department of Decision Sciences and IGIER. Email: \texttt{satoshi.fukuda@unibocconi.it}.} \and Nenad Kos\thanks{Bocconi University, Department of Economics, IGIER and CEPR. Email: \texttt{nenad.kos@unibocconi.it}.}}
\date{\today}
\begin{document}

\maketitle
\thispagestyle{empty}

\begin{abstract}
We study a behavioral SIR model with time-varying costs of distancing. The two main causes of the variation in the cost of distancing we explore are distancing fatigue and public policies (lockdowns). We show that for a second wave of an epidemic to arise, a steep increase in distancing cost is necessary. Distancing fatigue cannot increase the distancing cost sufficiently fast to create a second wave. However, public policies that discontinuously affect the distancing cost can create a second wave. With that in mind, we characterize the largest change in the distancing cost (due to, for example, lifting a public policy) that will not cause a second wave. Finally, we provide a numerical analysis of public policies under distancing fatigue and show that a strict lockdown at the beginning of an epidemic (as, for example, recently in China) can lead to unintended adverse consequences. When the policy is lifted the disease spreads very fast due to the accumulated distancing fatigue of the individuals causing high prevalence levels. 

\vspace{.5cm}
\end{abstract}

\clearpage
\pagenumbering{arabic}
\setcounter{page}{1}

\section{Introduction}

During an epidemic, the health and possibly even the lives of individuals are at risk. Each interaction with another (possibly contagious) individual leads to the possibility of infection. As a protective measure, people tend to limit their social interactions. Two primary factors influence individuals' protective behavior: the probability of getting infected and the cost associated with abstaining from social interactions. This paper explores how fluctuations in the distancing cost shape individuals' endogenous distancing behavior and how this affects the dynamics of an epidemic.

The reasons for variations in distancing cost are numerous. For example, distancing fatigue leads to a gradual increase in distancing cost as individuals deprive themselves of social interaction.\footnote{The World Health Organization Regional Office for Europe defines ``pandemic fatigue'' as demotivation to follow recommended protective behaviors, emerging gradually over time and affected by several emotions, experiences, and perceptions as well as the cultural, social, structural, and legislative environment \citep{WHO_Europe_20}.} \citet{Franzen_Wohner_21} document distancing fatigue among young adults in Switzerland during the COVID-19 pandemic.\footnote{There is growing evidence that distancing fatigue reduces the effectiveness of a mitigation policy. See, for instance,  \citet*{Goldstein_et_al_21}, \citet{joshi2021lockdowns}, \citet{Petherick_et_al_2021} and \citet{Du_et_al_2022}.} On a more basic level, there is a long line of research documenting how social groups increase the well-being of individuals by offering safety and increased odds of survival.\footnote{See, for instance, \citet{harlow1959affectional}, \citet{bowlby1969attachment}, \citet{baumeister1995need}, and \citet{Eisenberger_12}. \citet{Matthews_et_al_2016} show that after 24 hours of isolation mice search for social interaction and dopamine neurons in mice brain show similar patterns of activation as in other cravings.} Moreover, religious festivals such as Christmas and Holi or seasonal festivals such as Thanksgiving and the Chinese New Year make it more difficult for people to avoid social interactions and, thus, correspond to a sudden and short-term rise in distancing cost.\footnote{According to the American Automobile Association, nearly 56 million people traveled during the 2019 Thanksgiving (https://newsroom.aaa.com/2022/11/thanksgiving-travel-ticks-up-just-shy-of-pre-pandemic-levels/). The Chinese New Year may have been ``the biggest human migration on the planet'' (https://edition.cnn.com/travel/article/lunar-new-year-travel-rush-2019/index.html), at least until 2019.} Government policies enacted during an epidemic effectively decrease the distancing cost. For example, closures of restaurants and movie theaters reduce the availability of activities with individuals interacting and thereby encourage social distancing. Conversely, the lifting of such a policy increases the distancing cost. \citet{Hatchett7582}, \citet{Bootsma_Ferguson_07}, and \citet{caley2008} demonstrate that relaxations in non-pharmaceutical interventions increased social activity during the 1918 influenza pandemic. \citet{nguyen2020impacts} finds an increase in mobility soon after US-states reopened during the COVID-19 pandemic.


Motivated by the empirical work outlined above, we explore an SIR epidemiological model in which myopic individuals choose how much to distance at each point in time while the distancing cost may change over time. Distancing of myopic individuals is studied in \citet{Dasaratha_20},  \citet{Avery_2021}, \citet{engle2021behavioral}, \citet*{McAdams_Song_Zou_23}, and \citet*{Carnehl_Fukuda_Kos_23}.\footnote{To the best of our knowledge, there is no analytical characterization of equilibrium distancing behavior in an SIR model with far-sighted individuals  even when distancing costs are fixed over time. Nevertheless, we present the results of numerical simulations with farsighted individuals that support our main theoretical insights.} These papers show that with a constant cost of distancing, the prevalence is single-peaked. In this paper, we show that for a second peak of an epidemic to arise, the cost of distancing would have to rise extremely rapidly at some point after the first peak. We then proceed to analyze two main applications: distancing fatigue and public policies. 

In a framework with distancing fatigue, the cost of distancing increases in the discounted amount of past  distancing; for an axiomatization, see \citet*{Baucells_Zhao_19}. We show that even under distancing fatigue the prevalence has a single peak. After the prevalence peaks first, the growth of fatigue slows down up to the point at which fatigue starts decreasing. As a consequence, a second wave of the infection cannot arise from distancing fatigue alone. This suggests that the distancing cost is also single-peaked and that it peaks at least as late as prevalence. Yet, quantitatively, distancing fatigue heightens peak prevalence potentially burdening the health care system.

While distancing fatigue cannot cause a second wave of an epidemic, a sharp, sudden rise in the cost of distancing can. Public holidays and festivities---when it becomes challenging for individuals to keep social interactions low---or the termination of a mitigation policy---when the distancing cost discretely rises---can generate such sharp increases. To better understand these increases, we characterize a threshold distancing cost function: by how much would, at each point in time, the distancing cost have to increase or decrease instantaneously to change the sign of the slope of prevalence. 

The threshold distancing cost is particularly useful for two purposes. First,  when the prevalence is increasing, our characterization shows how much the cost of distancing would have to fall for the prevalence to start decreasing. This information is crucial for a policymaker weighing the harshness of non-pharmaceutical interventions in an attempt to reverse the course of an epidemic. Second, when the prevalence is decreasing, it determines the largest amount by which the distancing cost could increase without causing a second wave, thereby providing vital information for a policymaker considering to lifting a mitigation policy. If the policymaker bases their decisions to lift policies based on current prevalence and immunity alone while ignoring distancing fatigue, an unintended second wave may arise.

In addition, we illustrate that our equilibrium model with a time-varying distancing cost can be useful for other purposes. Several papers have analyzed optimal mitigation policies in a reduced form by assuming that a planner directly controls the transmission rate of a disease over time.\footnote{See the literature on the macroeconomic costs of an epidemic and the one on an optimal control of an epidemic: among others, \citet{Gonzalez-Eiras_Niepelt_20}, \citet*{Hall_Jones_Klenow_20}, \citet{acemoglu2020multi}, \citet*{alvarez2021simple}, \citet*{Eichenbaum_et_al_21}, \citet*{Farboodi_2021_internal},  and \citet{Kruse_Strack_22}.} However, they do not explicitly incorporate individuals' endogenous responses to such policies and how these transmission rates would be implemented. We show how different, time-varying transmission rates can be implemented in equilibrium with endogenous distancing by controlling the distancing cost, such as closing restaurants or restricting the occupation rate of public indoor places. This analysis reveals that papers studying optimal mitigation should consider alternative cost functions of transmission reduction taking prevalence into account. When the prevalence is high, endogenous distancing already leads to a substantial reduction in the transmission rate. 


Finally,  we conduct a numerical analysis on the impact of a stringent lockdown during the initial phase of an epidemic, drawing parallels with the COVID-19 lockdown in China. We find that such a lockdown can lead to a greater total number of infections compared to a scenario without a lockdown and that it can cause a second wave of the epidemic. The lockdown at the beginning of an epidemic postpones the spread of the infection (unless it eradicates it). Once it is lifted, individuals have accumulated a substantial level of distancing fatigue. As a consequence, lifting the lockdown leads to less endogenous protective distancing and thus more social interactions than without a lockdown. Thus, the combination of distancing fatigue and the sharp rise in distancing cost once the lockdown is lifted, can lead to a substantial second wave of the epidemic causing a greater total amount of infections.

\textbf{Related Literature.} To the best of our knowledge, this is the first paper to comprehensively study the effects of a time-varying cost of distancing in an SIR model with behavior. However, our model builds on work developed over the last century. We do our best to give credit to these foundations and other related work. 

The building blocks of the SIR model were set by the seminal work of \citet{Ross_Hudson_1917} and \citet{Kermack_McKendrick_27}. The incorporation of preventive behavior in such models, however, is a more recent endeavor. \citet{Reluga_10}, \citet{Fenichel_et_al_11}, \citet{Chen_12}, and \citet{Fenichel_13} introduced social distancing into SIR models and provided numerical analyses of equilibrium trajectories.  

Models of distancing with myopic agents, are analyzed by \citet{Dasaratha_20},  \citet{Avery_2021}, \citet{engle2021behavioral}, \citet*{McAdams_Song_Zou_23}, and \citet*{Carnehl_Fukuda_Kos_23}.\footnote{\citet{Rachel_20_Analytical} and \citet{Toxvaerd_20} analyze the model with non-myopic agents and offer two sets of results that contradict each other. Both arguments that the behavior under analysis is equilibrium behavior are incomplete.} 
 The last two papers establish the single-peakedness of equilibrium prevalence. \citet{Dasaratha_20} analyzes a model where the individuals are uncertain whether they are infected. \citet{Avery_2021} studies the interplay between distancing behavior and the willingness to get vaccinated. In relation to distancing fatigue, \citet{Avery_2021} models fatigue as an increase in the cost of distancing after a certain amount of time, independently of the previous amount of distancing. His analysis revolves around the effects of fatigue on the adoption of vaccines. In \citet*{McAdams_Song_Zou_23}, each individual's distancing cost varies over time because it depends on other non-infected individuals' distancing. \citet{engle2021behavioral} propose a behavioral SIR model with myopic agents but with a different meeting rate. 

Papers such as \citet{Brett_Rohani_20}, \citet{Gualtieri_Hecht_21}, \citet*{MacDonald_Browne_Gulbudak_21}, and \citet{Meacci_Primicerio_21} have proposed non-behavioral SIR models to study the effect of epidemic fatigue on the dynamics of an epidemic. Roughly, such non-behavioral SIR models introduce a new compartment that corresponds to epidemic fatigue (e.g., a new susceptible compartment with a higher transmission rate  due to epidemic fatigue). Our contribution is to study distancing fatigue within a behavioral SIR model. 

Other papers have studied possibilities and reasons behind second waves. \cite{Rachel_20_Second} studies the likelihood of a second wave of an epidemic in a behavioral SIR model and argues that lifting a mitigation policy can lead to a second wave if the society has not achieved herd immunity. Our paper sheds light on the effect of endogenous social distancing on the resurgence of an epidemic through the threshold distancing cost function. We show that even partially lifting a mitigation policy may trigger a second wave as individuals have more incentives to expose themselves after the policy is lifted (in addition to the fact that there are more susceptibles). Distancing fatigue exacerbates this issue. Numerical projections for the COVID-19 pandemic in \citet*{Giannitsarou_et_al_2021} suggest that waning immunity can cause several waves. Our result that distancing fatigue alone does not cause a second wave complements their finding.



\section{Model}

We study distancing behavior of individuals whose distancing cost varies over time and how this affects dynamics of an epidemic. We do so within the most prominent epidemic model, the SIR model.

A continuum of individuals, indexed by $i \in [0,1]$, is infinitely lived with time labeled by $t \in [0, \infty)$. The population is divided into three compartments: susceptible ($S$), infected ($I$) and recovered ($R$).\footnote{As the first paper that looks at the effect of time-varying distancing cost on equilibrium social distancing, our model abstracts from various features such as deaths and vaccines.} Susceptible individuals can get infected by meeting an infected individual. Infected individuals recover at rate $\gamma>0$. This implies that it takes on average $1/\gamma$ units of time to recover. After recovery, individuals acquire permanent immunity and cannot get infected again.\footnote{We assume that individuals know in which state they are. \cite{Dasaratha_20} and \cite{baril2021self} study environments where individuals are uncertain of the state they are in.} The size of the population is constant over time: $S(t) + I(t) + R(t) = 1$ for all $t \geq 0$. 

Individuals are responsive to the threat of infection and thus might try to avoid it. We capture this by letting a susceptible individual $i$ choose the level of exposure to the infection $\varepsilon_{i}(t) \in [0,1]$ at each point in time. The susceptible individual who chooses exposure $\varepsilon_{i}(t)$ at time $t$ gets infected at rate $\beta \varepsilon_{i}(t) I(t)$, where $\beta>\gamma$ is the transmission rate of the disease. Less exposure, i.e., lower $\varepsilon_{i}(t)$, thus, decreases the chance of infection. In the absence of the epidemic, the individual would go about her daily business with $\varepsilon_{i}(t)=1$. Conversely, we define $i$'s distancing at time $t$ as $d_{i}(t) := 1- \varepsilon_{i}(t)$. We assume that getting infected comes at a cost $\eta \geq 0$ while being susceptible generates a flow payoff of $\pi_{S}$. The assumption that the cost of infection is constant over time is akin to assuming that the individuals are myopic.\footnote{This approach has been frequently adopted in the recent theoretical literature on equilibrium social distancing. See, for example, \citet{Dasaratha_20},  \citet{Avery_2021}, \citet{engle2021behavioral}, \citet*{McAdams_Song_Zou_23}, and \citet*{Carnehl_Fukuda_Kos_23}. In contrast, in the model with farsighted individuals, the cost of infection $\eta$ serves as a co-state variable (to the probability of being susceptible), which varies over time. The difficulty in analyzing such a case comes from the fact that the co-state variable $\eta$ is a forward-looking variable that depends on the entire future paths of behaviors and the disease. We formally illustrate this point in Appendix \ref{sec:far_sighted}, and we confirm that our main insights hold for far-sighted individuals through numerical simulations.} The standard non-behavioral SIR model corresponds to the case with $\eta=0$. A reduction in exposure comes at a cost $\frac{c_{i}}{2}(t)(1-\varepsilon_{i}(t))^{2}$. The main novelty of our model is that we allow individual $i$'s \textit{distancing cost} $c_{i}(t)$ to vary over time. 

More precisely, for each susceptible individual $i$, the distancing cost is a piece-wise continuously differentiable function $c_{i}: [0, \infty) \rightarrow [\underline{c}, \infty)$ with the following three properties: (i) there exists a lower bound $\underline{c} > 0$ such that $c_{i}(t) \geq \underline{c}$ for all $t$; (ii) there are at most a finite number of jump discontinuities of $c_{i}$, which are common for all individuals $i$, at $t_{1} < \dots < t_{N}$ such that, on each interval $(t_{n}, t_{n+1})$ with $n \in \{ 1, \dots, N\}$,\footnote{For ease of exposition, let $t_{N+1}=\infty$.} $\dot{c}_{i}(t)$ is a continuous function satisfying
\begin{equation}\label{eq:distancing_cost_defi}
\dot{c}_{i}(t) = F(t, c_{i}(t), d_{i}(t) ),
\end{equation}
where $F(t, \cdot, \cdot)$ is a function of $i$'s current distancing cost $c_{i}(t)$ and her current distancing level $d_{i}(t)$;\footnote{To focus on the effect of distancing fatigue (i.e., the effect of current distancing on future distancing costs), we do not consider the case in which an individual's distancing cost depends on the distancing levels of the other individuals. For such a peer effect, see \citet*{McAdams_Song_Zou_23}.} and (iii) at each $t_{n}$ with $n \in \{ 1, \dots, N \}$, $c_{i}$ is right-continuous. At $t=0$ and at any point $t$ of jump discontinuity of $c_{i}$, the value of $c_{i}(t)$ is exogenously given. Discrete jumps allow to accommodate holidays (in which the distancing costs discretely jump) or changes in social-distancing policies (that also discretely affect the distancing costs). In addition, we assume that $c_{i}(0)$ is independent of $i$ and denote it $c_{0}$. A clarification is in order. While $c_i(t)$ may depend on the identity of an individual $i$, the environment is symmetric due to the common law of motion $F$ and the common initial cost $c_0$. Differences in the cost among individuals might, however, arise due to variations in the choice of distancing.

\subsection{Main Applications}\label{ssec:time_varying} 
Two forms of time-varying distancing cost are of particular interest: (i) distancing fatigue, that is, the decline in individuals' willingness to reduce their social activities to prevent infections, and (ii) policy interventions, such as restaurant closures and lockdowns.

We model distancing fatigue by having individuals' cost of distancing depend cumulatively on all the previous distancing decisions
\begin{equation}\label{eq:c_distancing_0}
c_{i}(t) = c_{0} + \underbrace{k\int_{0}^{t} e^{-r(t-\tau)} (1-\varepsilon_{i}(\tau))d\tau}_{=: \varphi_{i}(t)},
\end{equation}
where $k \geq 0$ and $r > 0$ are constants; details follow in Section \ref{sec:continuous_costs}. Thus, an individual's distancing cost $c_{i}(t)$ depends on a baseline distancing cost $c_0$ and the current level of distancing fatigue $\varphi_{i}(t)$. 

The above function captures two important properties of fatigue. First, past distancing increases each individual's distancing cost. The constant $k$\textemdash the fatigue accumulation rate\textemdash captures the rate at which current distancing increases the distancing cost. Second, the effect of past distancing choices on the cost of distancing decays over time at the fatigue recovery rate $r$. \citet*{Baucells_Zhao_19} provide a decision-theoretic axiomatization of the fatigue utility model of this form. 

Second, Section \ref{sec:discountinuous_costs} models policy interventions as a reduction in the distancing cost $c_{i}(t)$. When restaurants are closed and/or in-person activities are constrained, the opportunity cost of distancing decreases. In particular, we consider a time-varying policy variable $\ell(t) \in [\overline{c}/c_0,1]$ with at most finitely many discontinuities such that 
\begin{equation*}
c_{i}(t) = c_0 \cdot \ell(t) \in [\bar{c}, c_0],
\end{equation*}
that is, $\ell(t)$ can be interpreted as the strictness of the policy intervention at time $t$.\footnote{Note that we could also accommodate periods of increased distancing cost, such as holidays, straightforwardly by allowing $\ell(t)>1$.}

Finally, Section \ref{sec:lockdown_and_fatigue} discusses the interaction between these two time-varying distancing costs, that is, a model with policy interventions and distancing fatigue. In that case, the distancing cost function will be 
\begin{equation*}
c_{i}(t) = c_0 \cdot \ell (t) + \varphi_{i}(t).
\end{equation*}

\subsection{Equilibrium and Behavior}\label{ssec:equilibrium}
This subsection defines an equilibrium of our model, proves its existence and uniqueness, and characterizes individuals' equilibrium level of distancing as a function of the current distancing cost and the state of the epidemic.

A susceptible individual $i$ determines her current exposure level by solving:
\begin{equation}\label{eq:naive_prob}
\max_{\varepsilon_{i}(t) \in [0,1]} \pi_{S} - \frac{c_{i}(t)}{2} (1-\varepsilon_{i}(t))^{2} - \beta \eta I(t) \varepsilon_{i}(t) .
\end{equation}
At each time $t$, the susceptible individual $i$ takes the value of $c_{i}(t)$ as given, while the resulting exposure level affects the slope of the distancing cost, $\dot{c}_{i}(t)$.\footnote{Intuitively, consider a discrete-time model in which, at the start of each period, a susceptible individual takes her distancing cost at that time as given. Her resulting exposure level affects her distancing cost at the beginning of the next period. Our model would correspond to the continuous-time limit of such a model.}

Let the average exposure be $\varepsilon(t) := \frac{1}{S(t)} \int_{} \varepsilon_i(t) di$, where the integral is taken over the susceptible individuals. The disease dynamics are governed by the following system of differential equations:
\begin{align}
\dot{S}(t) &=-\beta \varepsilon(t) I(t) S(t), \label{eq:S_dot} \\
\dot{I}(t) &= I(t)(\beta \varepsilon(t)S(t)-\gamma), \label{eq:I_dot} \\
\dot{R}(t) &= \gamma I(t), \label{eq:R_dot}
\end{align}
for all except possibly a finite number of $t$, with the initial condition $(S(0),I(0),R(0)) = (S_{0}, I_{0},0)$ with $I_{0} \in (0,1)$ and $S_0 = 1-I_0$. With these in mind, we define an equilibrium.

\begin{defi}
An \textit{equilibrium} is a tuple of functions $( S, I, R,  (c_{i}, \varepsilon_{i})_{i})$ with the following three properties: (i) $(S, I, R)$ are continuous functions that satisfy (\ref{eq:S_dot}), (\ref{eq:I_dot}) and (\ref{eq:R_dot}) with the initial condition $(S(0),I(0),R(0)) = (S_{0},I_{0},0)$, where $\varepsilon$ is the average exposure;\footnote{The assumption of continuity of $(S,I,R)$ is innocuous in light of our application. Discontinuities could only arise at discontinuities of $c(t)$. However, $c(t)$ only affects behavior and the changes in $I$ and $S$ over an interval of time of length $\Delta>0$ is bounded from above by the SIR dynamics without behavior and from below by a path in which no individual is infected in this interval. Both paths are continuous as $\Delta$ goes to zero and so are our modified dynamics.} (ii) each $\varepsilon_{i}$ solves (\ref{eq:naive_prob}), that is, $\varepsilon_{i}$ is a best response to $(S,I,R)$ given $c_{i}$; and (iii) the distancing cost function $c_{i}$ satisfies (\ref{eq:distancing_cost_defi}), where $d_{i} = 1-\varepsilon_{i}$. An equilibrium is \textit{symmetric} if $\varepsilon = \varepsilon_{i}$ for all $i$. 
\end{defi}

As the susceptible individual's objective function is concave in her exposure level, the first-order condition of the individual's problem yields 
\begin{equation}\label{eq:naive_distancing}
\varepsilon_{i}(t)= \max \left( 0, 1 - \frac{\beta \eta I(t)}{c_{i}(t)} \right).
\end{equation}

An individual chooses a lower exposure (that is, she distances more) when the prevalence is higher. Distancing increases in the cost of infection $\eta$ and the transmission rate $\beta$, and decreasing in the cost of distancing $c_i(t)$. In equilibrium, $\varepsilon_i = \varepsilon$ and $c:=c_i$ for all $i$ due to \eqref{eq:naive_distancing} differing across individuals only in the distancing cost $c_i(t)$ and the fact that $c_i(0)=c_0$ for all $i$. Therefore, any equilibrium is symmetric:
\begin{equation} \label{eq:naive_distancing_symmetric}
\varepsilon(t)=\max \left( 0, 1 - \frac{\beta \eta I(t)}{c(t)} \right).
\end{equation}
Plugging the expression for exposure (\ref{eq:naive_distancing_symmetric}) into the system of differential equations (\ref{eq:S_dot}), (\ref{eq:I_dot}), (\ref{eq:R_dot}), and (\ref{eq:distancing_cost_defi}) leads to the system of differential equations characterizing the equilibrium. We denote by $(S,I,R,c,\varepsilon)$ the symmetric unique equilibrium. Summarizing our discussions, we obtain:

\begin{prop}\label{prop:existence}
An equilibrium exists, is unique and symmetric. In the unique equilibrium, the system $(S,I,R)$ satisfies $\displaystyle I_{\infty} := \lim_{t \rightarrow \infty} I(t) = 0$, $\displaystyle S_{\infty} := \lim_{t \rightarrow \infty} S(t) \in \left( 0, \frac{\gamma}{\beta} \right)$, and $\displaystyle \lim_{t \rightarrow \infty} \varepsilon(t) =1$. 
\end{prop}

In our environment, the prevalence disappears in the limit as time goes to infinity, and individuals return to full exposure. The final size of susceptibles, $S_{\infty}$, is below the threshold of herd immunity $\frac{\gamma}{\beta}$.  


Throughout most of the analysis, we focus on the case in which the prevalence is increasing at the outset: $\dot{I}(0) > 0$. Substituting (\ref{eq:naive_distancing_symmetric}) into (\ref{eq:I_dot}) at time $t=0$, this occurs whenever
\begin{equation}\label{eq:I_taking_off}
\beta S_{0} \left( 1- \frac{\beta \eta I_{0}}{c_{0}} \right) - \gamma >0.
\end{equation}
The above inequality is satisfied as long as $\beta>\gamma$ and the initial seed of infection $I_0$ is small enough.\footnote{Alternatively, given the fixed values of parameters other than $\beta$, there exist $\underline{\beta}$ and $\overline{\beta}$, with $\gamma < \underline{\beta} < \overline{\beta}$, such that $\dot{I}(0) > 0$ if and only if $\beta \in (\underline{\beta}, \overline{\beta})$.}



\section{Continuous Distancing Cost and Distancing Fatigue}\label{sec:continuous_costs}

This section studies the case in which the distancing cost is a continuous function of time. Section \ref{sec:continuous_cost_time} provides a general sufficient condition for an epidemic to be single-peaked for a very general specification of the distancing cost. Section \ref{sec:continuous_fatigue} investigates the model with distancing fatigue. We show that the equilibrium prevalence peaks at most once and discuss some consequences of distancing fatigue.

\subsection{Sufficient Conditions for Single-Peaked Epidemics}\label{sec:continuous_cost_time}

A single peak of an epidemic is one of the most prominent qualitative features of the SIR model; see, for example, \citet{brauer2012mathematical}. Here, we examine conditions on the distancing cost such that this feature remains intact. Conversely, this provides intuition for the properties of the time-varying distancing cost that are required for a second wave to arise. In the following, we define peak prevalence as a strict local maximum. 
 


\begin{prop}\label{prop:condition_one_wave}
Let $c$ be a (continuously) differentiable function such that $\dot{c}$ is given by (\ref{eq:distancing_cost_defi}) for all $t >0$. If
\begin{equation}\label{eq:condition_one_wave}
\frac{\dot{c}(t)}{c^{2}(t)} < \frac{\varepsilon^{2}(t)}{\eta}  \text{ for all } t >0, 
\end{equation}
then, in equilibrium, prevalence $I$ has a single peak. A sufficient condition for the above inequality to be satisfied is
\begin{equation}\label{eq:condition_one_wave_2}
\frac{\dot{c}(t)}{c^{2}(t)} <  \frac{1}{\eta} \left( \frac{\gamma}{\beta S_0} \right)^{2} \text{ for all } t >0.
\end{equation}
\end{prop}



If the distancing cost is growing slowly, that is, if $\dot{c}/c^{2}$ is small, then $I$ has only one local maximum. The prevalence either decreases immediately and never picks up or increases from the outset until it reaches the peak and decreases thereafter. The proof argues that for a second wave to arise, the prevalence would first need to attain a local minimum at some time $t>0$ for which to occur $\dot{I}(t) =0$ and $\ddot{I}(t) \geq 0$ are necessary. By taking the derivative of $\dot{I}$, the latter requirement can be shown to be equivalent to 
\begin{align*}
\frac{\dot{c}(t)}{c^{2}(t)} \geq \frac{\varepsilon^{2}(t)}{\eta}.
\end{align*}
Therefore, if $\dot{c}(t)/c^2(t) < \varepsilon^{2}(t)/\eta$ for all $t>0$, there can be no local minimum and consequently no second wave. This result nests the time-invariant distancing cost $c(t) =c_{0}$ as a special case, an environment, which was previously analyzed in \citet*{Carnehl_Fukuda_Kos_23}, and the non-behavioral SIR model in the sense that $\eta=0$ implies that the right-hand side of (\ref{eq:condition_one_wave}) is infinity.


Moreover, note that if the reverse condition $\dot{c}(t)/c^2(t) > \varepsilon^{2}(t)/\eta$ would hold for all $t>0$, then any stationary point of $I$ was a local minimum. This contradicts the existence of a differentiable cost function such that $\dot I(0)>0$, as $I_{\infty} =0$. Indeed, a similar result can be established directly. Consider the inequality $\dot{c}(t)/c^2(t) > 1/\eta$, which is sufficient for the above inequality and let 
\begin{align*}
\sigma (t) := \frac{\dot c(t)}{c(t)}
\end{align*}
be the semi-elasticity of $c$. The above inequality can be rewritten as $\sigma (t) > c(t)/\eta$, that is, the semi-elasticity of a function dominates the function itself (normalized by positive constant $\eta$).

The next result establishes that no differentiable function exists that is everywhere positive and that satisfies this property.

\begin{lmm}\label{lmm:continuous_c_sufficient}
There does not exist a continuously differentiable function $c: [0,\infty) \rightarrow [\underline{c}, \infty)$ with $c(t)\geq \underline{c}$ for all $t$, such that  $\sigma(t) > \frac{c(t)}{\eta}$ for all $t>0$.
\end{lmm}

The inequality $\sigma(t) > c(t)/\eta$, which is necessary for a second wave to arise, requires the distancing cost $c$ to grow extremely fast. To build further intuition, fix $c_{0}>0$. A solution to the differential equation $\sigma(t) = c(t)/\eta$ on $t \in \left( 0, \eta/c_{0} \right)$ is
\begin{equation*}
c(t) = \frac{1}{\frac{1}{c_{0}} - \frac{t}{\eta}},
\end{equation*}
which grows towards infinity as $t$ goes towards $\eta/c_{0}$. 

This analysis establishes that a continuous distancing cost function that generates more than one wave needs to satisfy $\dot{c}(t)/c^2(t) < \varepsilon^{2}(t)/\eta$ for low values of $t$. More precisely it must do so at the first stationary point, followed by a period where $\dot{c}(t)/c^2(t) > \varepsilon^{2}(t)/\eta$. The second period must, however, be limited in duration as otherwise there does not exist a distancing cost function that satisfies these conditions. That is, for the second wave to arise there needs to be an abrupt continuous change in the distancing cost or a discontinuity after a first peak. In Appendix \ref{sec:dec_at_0} we apply Proposition \ref{prop:condition_one_wave} to various distancing-cost functions.

The following subsection applies the above analysis to distancing fatigue and shows that the prevalence is single-peaked.

\subsection{Distancing Fatigue}\label{sec:continuous_fatigue}

It is in human nature to socialize, and maintaining social distancing over a long time horizon becomes increasingly difficult. This phenomenon, distancing fatigue, and its importance have been well-documented empirically during the COVID-19 pandemic.\footnote{As discussed in the Introduction, see, for example, \citet{Franzen_Wohner_21}, \citet*{Goldstein_et_al_21}, \citet{joshi2021lockdowns}, \citet{Petherick_et_al_2021}, and \citet{Du_et_al_2022}.} 

As foreshadowed in Section \ref{ssec:time_varying}, we model distancing fatigue using the time-varying distancing cost function:
\begin{equation}\label{eq:c_distancing}
c(t) = c_{0} + k\int_{0}^{t} e^{-r(t-\tau)} (1-\varepsilon(\tau))d\tau,
\end{equation}
where $k \geq 0$ is the fatigue accumulation rate and $r > 0$ is the fatigue recovery rate.\footnote{When $k=0$, the distancing cost is constant over time. When $r=0$, there is no decay, and thus the cost of distancing is non-decreasing over time. While we rule out this case only for ease of exposition, the single-peakedness of the prevalence (Proposition \ref{prop:single_peak_discounting}) still holds under $r=0$.}  It follows that the fatigue 
\begin{equation*}
\varphi(t):= k\int_{0}^{t} e^{-r(t-\tau)} (1-\varepsilon(\tau))d\tau
\end{equation*}
increases in past distancing but the effect of past distancing on fatigue decays over time. 

In terms of the marginal change in the distancing cost, equation (\ref{eq:c_distancing}) is written as
\begin{equation}\label{eq:c_dot_discounting}
\dot{c}(t) =  k (1-\varepsilon(t)) - r (c(t)-c_{0}) , 
\end{equation}
with the initial condition $c(0)=c_{0}$.\footnote{One can interpret fatigue $c(t)-c_{0}$ as ``capital:'' $k$ is the saving rate, $1-\varepsilon(t)$ is the current production, and $r$ is the depreciation rate.} This differential equation is a special case of equation (\ref{eq:distancing_cost_defi}) and therefore the existence and uniqueness of equilibrium follow from Proposition \ref{prop:existence}.

\subsubsection{Single-Peaked Prevalence under Distancing Fatigue}

We start by developing two preliminary results. First, once individuals expose themselves to the disease to some extent, they will never fully distance themselves afterward (i.e., their exposure is strictly positive afterward). Second, we show that, at the moment at which the distancing cost attains a local maximum, prevalence is non-increasing. 

\begin{lmm}\label{lm:varepsilon_positive}
In equilibrium, if  $\varepsilon(t') >0$ for some $t'$, then $\varepsilon(t) >0$ for all $t \geq t'$. 
\end{lmm}

Intuitively, the result follows because exposure is inversely related to $I(t)/c(t)$. Therefore, should exposure fall to a very low level, $I(t)$ would become decreasing and the distancing cost increasing due to fatigue accumulating. These two effects would lead the exposure to grow, thereby preventing it from falling to $0$. In other words, $\varepsilon(t)$ can be equal to $0$ only at the outset of an epidemic. In that case, $\dot{I}(0) <0$. A simple assumption that guarantees $\varepsilon(t)$ to be strictly positive is $\varepsilon(0) = 1 - \frac{\beta \eta I_{0}}{c_{0}} >0$, which in turn is satisfied if $I_{0}$ is small enough.

Next, we show that any critical point of the distancing cost is directly related to the prevalence dynamics at the critical point. 

\begin{lmm}\label{lm:cI}
Suppose $c$ is given by (\ref{eq:c_distancing}) and $\dot{I}(0) >0$. In equilibrium, if $c$ attains a local maximum (minimum) at $t>0$, then $\dot{I}(t) \leq 0$ ($\geq 0$). 
\end{lmm}

If the distancing cost function has a local maximum, it must be at a time when the prevalence is declining. To see this, suppose that the maximum of the distancing cost function was attained when $I$ was increasing. At the same point, the exposure would decrease due to $c$ being locally flat and the prevalence increasing. However, then it cannot be that fatigue, and therefore the distancing cost function, is already maximized. This leads us to the main result of the section.

\begin{prop}\label{prop:single_peak_discounting}
Suppose $c$ is given by (\ref{eq:c_distancing}). If $\dot{I}(0)>0$, then the following hold: \vspace{-.6cm} \begin{enumerate}
\item The prevalence $I$ is single-peaked.
\item The distancing fatigue $\varphi$ and thus the distancing cost $c$ are single-peaked. 
\item The distancing fatigue $\varphi$ attains its peak no earlier than the prevalence $I$.
\end{enumerate}
\end{prop}

The first part of Proposition \ref{prop:single_peak_discounting} implies that distancing fatigue itself cannot cause a second wave. For a second peak to arise, the prevalence would have to attain a local minimum first and then begin to increase again. However, when the prevalence is falling, the growth of fatigue slows down and fatigue may even decrease. As a consequence, distancing is not decreasing fast enough to jump-start another wave. 

The single-peakedness of prevalence has important implications for other objects in the model as the second part of Proposition \ref{prop:single_peak_discounting} shows. 
First, since the fatigue recovery rate $r$ is strictly positive, fatigue is bounded from above:
\begin{equation}\label{eq:upper_bound_fatige}
\varphi(t) = k \int_{0}^{t} e^{-r(t-\tau)} (1-\varepsilon(\tau))d\tau \leq \frac{k}{r}.
\end{equation}
Second, the strictly positive fatigue recovery rate implies that fatigue cannot be strictly increasing in equilibrium over the entire time horizon. Since the cost of distancing is bounded from below by $c_0$, distancing will dissipate with the eventually vanishing prevalence. The fatigue recovery rate implies that, as time passes, fatigue will vanish too. 



\begin{figure}[t]
\begin{center}
\includegraphics[scale=0.9]{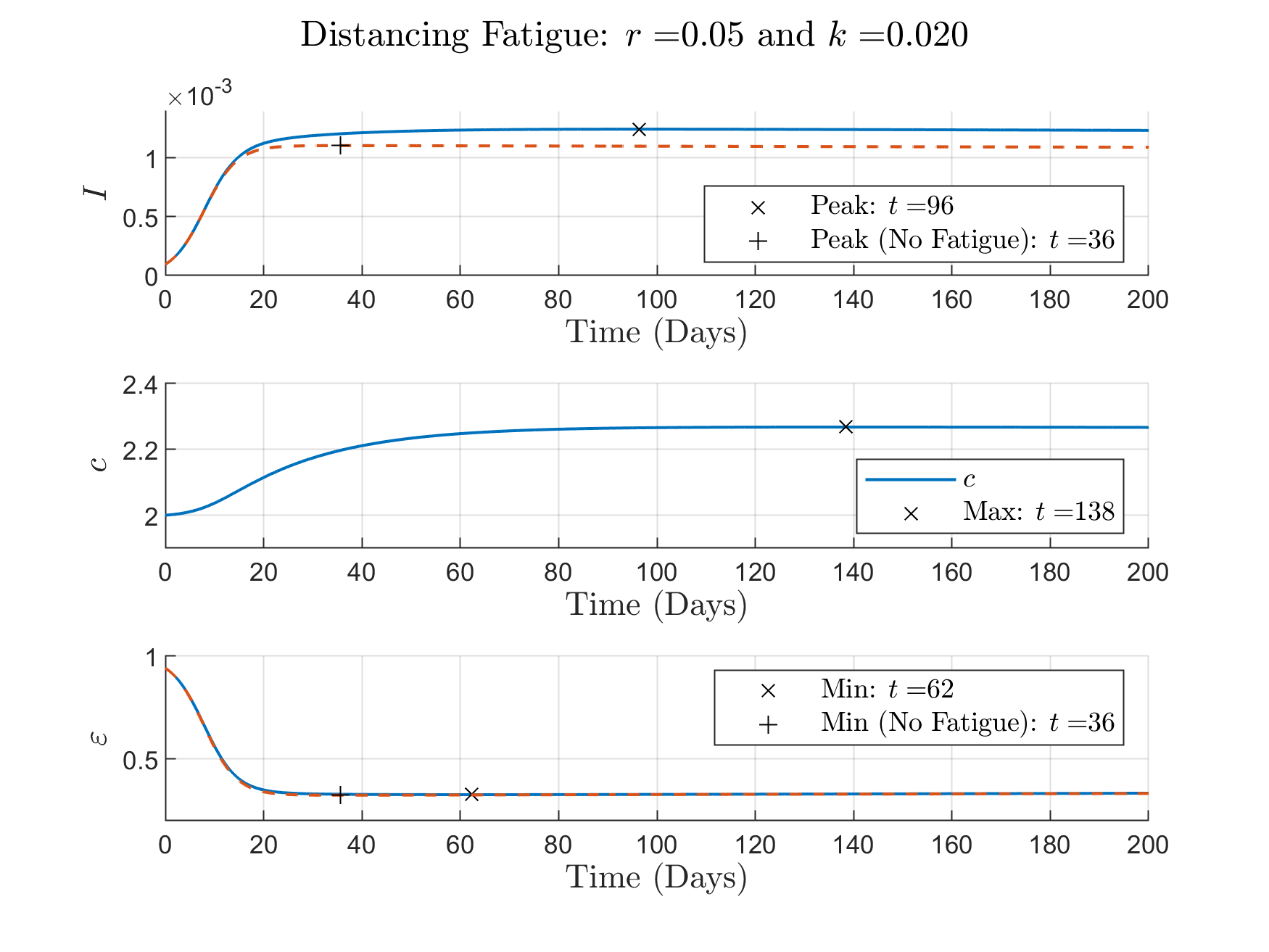} 
\end{center}
\caption{\emph{Distancing Fatigue}. The solid curves depict the prevalence $I$, the distancing cost $c$, and the exposure level $\varepsilon$ over time. For comparison, the dashed curves depict the analogous paths under constant distancing cost $c_{0}$.}\label{fig:fatigue_ex}
\end{figure}

Figure \ref{fig:fatigue_ex} illustrates Proposition \ref{prop:single_peak_discounting} 
when $c_{0}=2$, $r=0.05$, and $k=0.02$.\footnote{\label{fn:calibration}The other parameters are $(\beta, \gamma, I_{0}, \eta, c_{0}, k) = (0.3+\frac{1}{7}, \frac{1}{7},  0.95 \times 10^{-4}, 2761.63, 0.05, 0.005)$. The parameters $(\beta, \gamma, \eta)$ are calibrated for the onset of COVID-19 (when the cost of distancing is normalized at $c=2$) as in  \citet*{Carnehl_Fukuda_Kos_23}. Unless otherwise stated, we use these parameter values for $(\beta, \gamma, \eta)$ throughout the paper.} Prevalence initially increases relatively quickly which causes individuals to engage in more social distancing. Consequently, the distancing cost increases due to fatigue accumulating. As the cost increases and the prevalence approaches its peak, individuals start increasing their exposure level again. After some time, this causes fatigue, and thus, the distancing cost, to slowly decrease.\footnote{\label{fn:Rt1}Note that prevalence remains at an almost constant level in our model after the peak. This pattern has been empirically documented for the COVID-19 pandemic (e.g., \citealp*{atkeson2020four} and \citealp{Gans_22}). In fact, \citet{Gans_22} proposes a behavioral SIR model in which prevalence is \textit{assumed} to be constant to simplify the analysis.} 

Two remarks are in order. First, when $\dot{\varepsilon}(0) <0$, i.e., the susceptible individuals increase distancing at the onset of the epidemic, distancing peaks no later than the peak prevalence. This is because, when the prevalence peaks, the distancing cost is still increasing and thus distancing is decreasing. Since distancing is increasing at the onset of the epidemic, at some point no later than when the prevalence peaks, distancing has to peak. Figure \ref{fig:fatigue_ex} also illustrates this fact. Thus, there is a period (from day $62$ to $96$ in the figure) in which both prevalence and exposure increase.  


Second, as time goes to infinity, the distancing cost converges to its initial level: $\displaystyle\lim_{t \rightarrow \infty} c(t)=c_0$. In other words, individuals recover from fatigue, $\displaystyle\lim_{t \rightarrow \infty} \varphi(t)= 0 $. This is intuitive as the infection dies out and individuals stop distancing in the limit. 

To conclude, our findings suggest that distancing fatigue does not affect qualitative features of the prevalence trajectory. This, however, is not to say that distancing fatigue cannot play an important role in epidemiological models. 

First, it may very well have critical quantitative implications. \citet*{Goldstein_et_al_21}, for example, show that after four months of lockdown during the COVID-19 pandemic, non-pharmaceutical interventions had a significantly lower effect on reducing fatalities. In our model, the peak prevalence in the model with distancing fatigue is always higher than the peak prevalence in the model without distancing fatigue, and the peak prevalence in the model with distancing fatigue is reached no earlier than the one without distancing fatigue. Thus, distancing fatigue may burden the medical capacity constraint at the peak prevalence. 

Second, distancing fatigue introduces two opposing effects on individuals' distancing decisions. On the one hand, when the distancing cost increases due to distancing fatigue, ceteris paribus, the individuals increase their exposure because distancing becomes costlier. On the other hand, higher prevalence due to distancing fatigue makes it costlier for an individual to increase exposure. This second effect decreases individuals' exposure levels. Hence, to measure the effect of distancing fatigue on exposure, it is also important to measure the effect that an increased prevalence has on individuals' preventive behavior. 

Third, distancing fatigue introduces a negative dynamic spillover to lockdown policies. By encouraging or enforcing social distancing in the current period, the lockdown reduces distancing incentives in the future due to accumulated distancing fatigue. Holding lockdown stringency fixed, lockdown effectiveness declines over time and the likelihood of a second wave may increase should the lockdown be lifted. Our result that distancing fatigue alone does not cause the second wave is important because it suggests that the second wave may result rather from the discrete increase in distancing cost from lifting the lockdown policy. The next section studies the effect of a discrete change in distancing cost on the disease dynamics. Thereafter, we will investigate the interaction of lockdown policies with fatigue accumulation. 



\section{Discontinuous Distancing Cost and Policy Interventions}\label{sec:discountinuous_costs}

The cost of social distancing depends not only on previous exposure decisions but also on other factors, such as holidays or public health policies. The opportunity cost of social distancing sharply increases during holiday seasons when social gatherings have high value or during vacation times. \citet{mehta2021holiday} report an increase in travel and social activity during Thanksgiving 2020. \citet{Schlosser32883} document an increase in travel during school and public holidays. In contrast, business closures due to a governmental lockdown discretely lower the opportunity cost of social distancing. When public health policies are lifted or holidays pass, the cost of distancing returns to its initial level. 

The following subsection examines the effects of discontinuous changes in the cost of distancing. While the analysis to follow is cleanest with discontinuous changes, the results do not rely as much on the discontinuity as they do on sudden rapid changes in the cost of distancing. We focus on policy interventions that encourage social distancing behavior, that is, that reduce the distancing cost. However, introducing periods of increased distancing cost (i.e., holidays) can be straightforwardly implemented as well by allowing the distancing cost to increase. Appendix \ref{sec:holidays} analyzes such a case. Subsection \ref{sec:publicpoliciescost} shows how various ways of modeling public policies in the literature can be recast as changes in the distancing cost.

\subsection{Threshold Distancing Cost}
Before studying the effects of discontinuous changes in the distancing cost in detail, we introduce a useful technical tool to determine whether a policy change will lead to a second wave. In particular, we characterize the threshold on the distancing cost $\overline{c}(t)$ such that if $c(t)$ is above the threshold $\overline{c}(t)$, the slope of prevalence is positive, and if $c(t)$ is below the threshold $\overline{c}(t)$, the slope of prevalence is negative. The difference between the threshold and the actual distancing cost $c(t)$ is the largest instantaneous change in $c(t)$ that will not change the sign of the slope of $I(t)$. 


\begin{defi}\label{df:threshold_distancing_cost}
Let $c$ be a piece-wise continuously-differentiable distancing cost function and let $(S,I,R,c,\varepsilon)$ be the corresponding equilibrium. We define the threshold distancing cost function $\overline{c}$ as follows: for each $t \geq 0$,
\begin{equation*}
\overline{c}(t) := \begin{cases} 
\frac{\beta^2 I(t) S(t) \eta}{\beta S(t)-\gamma}, & \text{if } S(t)>\frac{\gamma}{\beta}  \\ \infty, & \text{if } S(t) \leq \frac{\gamma}{\beta} 
\end{cases}.
\end{equation*}
\end{defi}

In Definition \ref{df:threshold_distancing_cost}, note that $(S, I, R)$ are continuous functions satisfying (\ref{eq:S_dot}), (\ref{eq:I_dot}) and (\ref{eq:R_dot}) with the initial condition $(S(0),I(0),R(0)) = (S_{0},I_{0},0)$, where $\varepsilon$ is the average exposure that satisfies (\ref{eq:naive_distancing_symmetric}). The equilibrium is unique and symmetric. With this definition in mind:  

\begin{prop}\label{prop:threshold_cost}
Let $c$ be a piece-wise continuously-differentiable distancing cost function, and let $\overline{c}$ be the associated threshold distancing cost function. For any $t>0$ and piece-wise continuously-differentiable distancing cost function $c_2$ such that the corresponding equilibrium $(S_2,I_2,R_2,c_2,\varepsilon_2)$ satisfies the property that $c_2(s)=c(s)$ for all $s < t$, the following holds: 
\begin{align*}
\dot{I}_2(t_+):=\lim_{\tau \downarrow t} \dot{I}_2(\tau)<0\text{ if and only if }c_2(t) < \overline{c}(t). 
\end{align*}
\end{prop}

In words, the threshold distancing cost function $\overline{c}$ satisfies the following property. Fix a distancing cost function $c$ and its implied equilibrium, and consider an alternative distancing cost function $c_2$ that coincides with $c$ up to time $t$. Then, $\overline{c}(t)$ prescribes the largest value that the distancing cost $c_2(t)$ can take on such that the right-limit of the derivative of $I(t)$ under $c_2(t)$ is negative.\footnote{Note that $\overline{c}$ depends on the equilibrium path $(S,I, R, c, \varepsilon)$ under the distancing cost function $c$.}

Whenever $c$ is such that $I$ is single-peaked in equilibrium, the threshold distancing-cost function $\overline{c}$ intersects $c$ once and from below. In particular, as long as $\dot{I}(t) >0$, $\overline{c}(t) < c(t)$ and conversely so if $\dot{I}(t) >0$. In addition, when $S(t)$ approaches $\frac{\gamma}{\beta}$ from above, $\overline{c}(t)$ grows towards infinity. 

The difference $\overline{c} - c$ plays an important role. When the prevalence is decreasing, the cost difference informs by how much the cost can instantaneously increase without the prevalence starting to increase. Conversely, when the prevalence is already increasing, the difference $c-\overline{c}$ establishes by how much the cost of distancing must decrease for the prevalence to start falling. This is of particular interest to policymakers who are trying to establish the strictness of public health policies required to reduce the prevalence immediately. Conversely, it can be used to establish whether lifting a policy will lead to a second wave. \citet{Hatchett7582}, \citet{Bootsma_Ferguson_07}, and \citet{caley2008} suggest that, during the 1918 influenza pandemic, relaxations in non-pharmaceutical interventions caused a new surge of cases. 

In the remainder of this subsection, we demonstrate how the threshold function $\overline{c}$ evolves over time in two numerical examples. First, we consider a constant distancing cost function and illustrate how the threshold function $\overline{c}$ evolves in the absence of any changes. Second, we consider the introduction of a temporary lockdown and show how the threshold function $\overline{c}$ can guide policymakers.

\begin{figure}[t]
\begin{minipage}{0.49 \hsize}
\begin{center}
\includegraphics[scale=0.55]{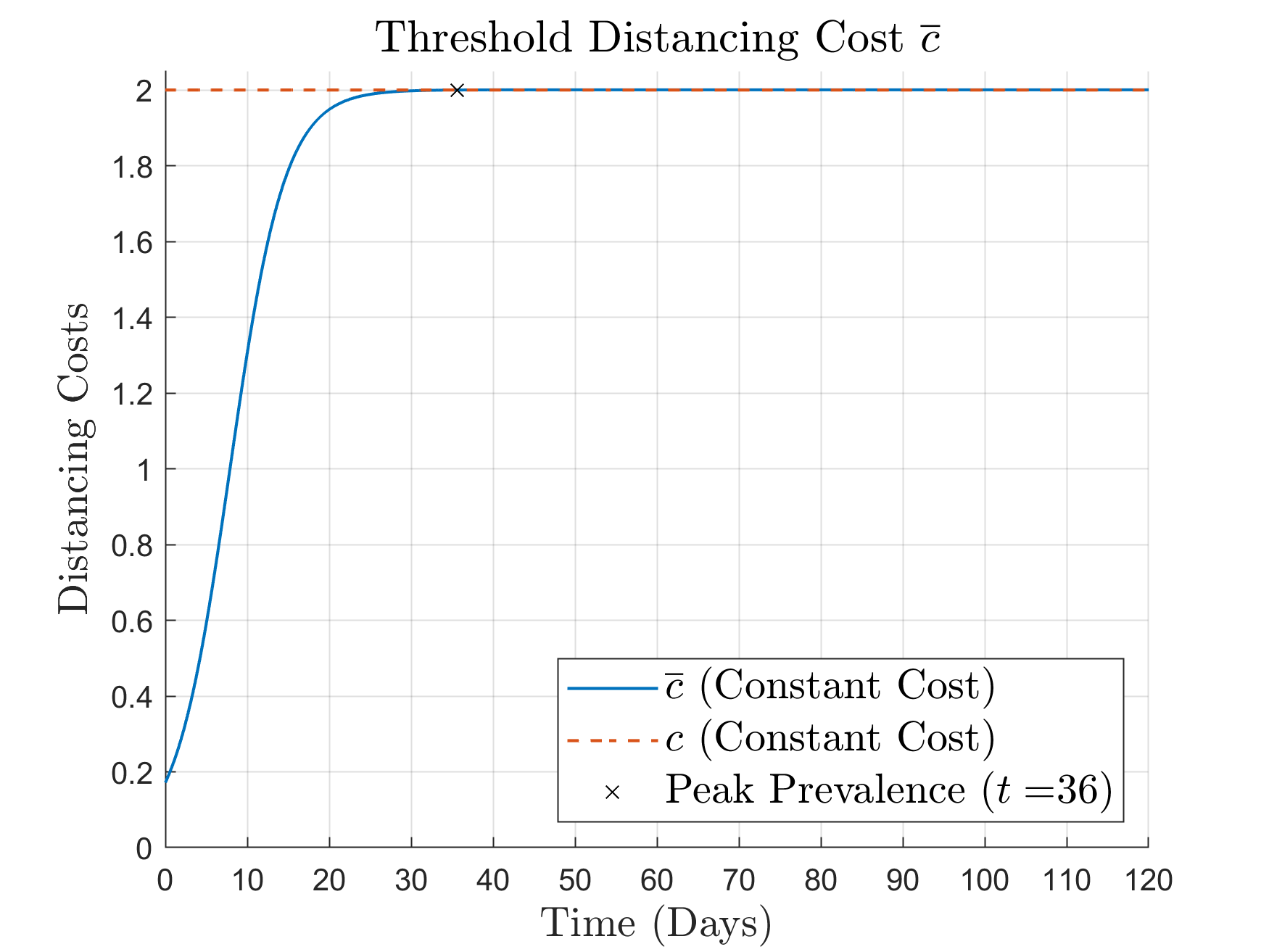}
\end{center}
\end{minipage}
\begin{minipage}{0.49 \hsize}
\begin{center}
\includegraphics[scale=0.55]{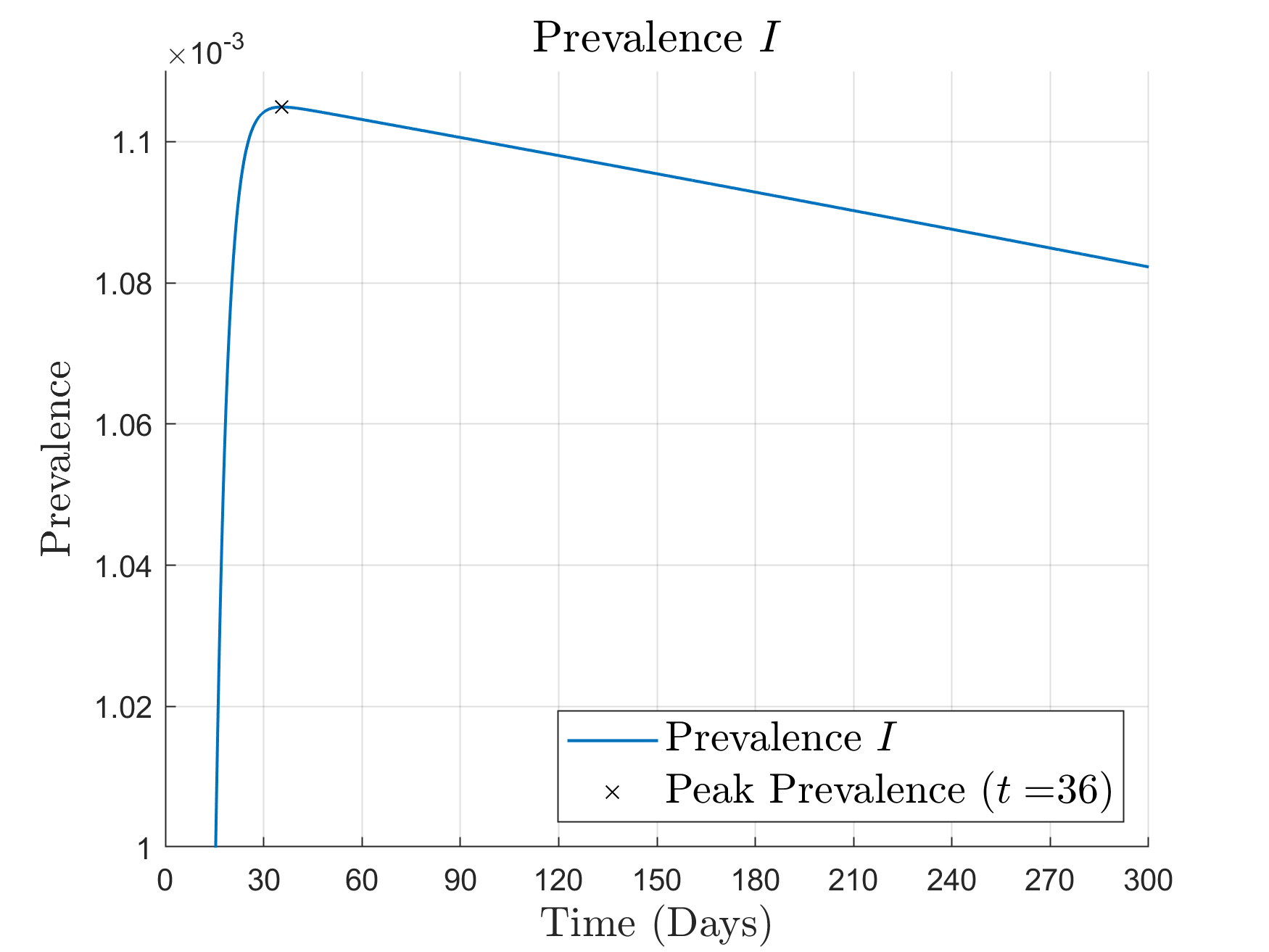}
\end{center}
\end{minipage}
\caption{\emph{Constant Distancing Cost}. The left panel depicts the threshold distancing cost function $\overline{c}$ over time. The right panel depicts the prevalence $I$ over time.}\label{fig:const_cost_c_bar}
\end{figure}

\begin{example}\label{exl:const_cost_c_bar}
To illustrate the threshold distancing cost $\overline{c}$, we consider a simple example in which the distancing cost $c$ is constant over time. The left panel of Figure \ref{fig:const_cost_c_bar} depicts the threshold distancing cost function $\overline{c}$ (solid curve) for the corresponding constant distancing cost function $c=2$ (dashed line). The right panel depicts the prevalence over time.\footnote{The example depicted in Figure \ref{fig:fatigue_ex} corresponds to the case in which distancing fatigue is added to the constant distancing cost case depicted in Figure \ref{fig:const_cost_c_bar}.} The peak prevalence is attained around day $36$.\footnote{While the prevalence theoretically declines towards $0$ in this constant distancing cost case, the right panel suggests that the decline is slow (recall footnote \ref{fn:Rt1}). The shape of the threshold distancing cost function $\overline{c}$ in the left panel of Figure \ref{fig:const_cost_c_bar} reflects this feature. Namely, $\overline{c}$, while growing, stays close to $2$ for an extended period of time after the peak. Although it is difficult to see from the figure, $c$ and $\overline{c}$ intersect only once (around day $36$).}
\end{example}

\bigskip
\begin{example}\label{exl:temporary_lockdown}
Next, we consider the introduction of a social-distancing policy. Letting the baseline distancing cost be $c(t)=c_0=2$, recall that the left panel of Figure \ref{fig:const_cost_c_bar} shows by how much the distancing cost $c$ must be reduced to decrease the prevalence before it would reach its peak otherwise. The threshold cost $\overline{c}(t)$ is $1.8$ around day 15 and the peak prevalence is attained on day 35. Suppose that the social-distancing measure $\ell(t) = 0.9$, which decreases distancing cost to $c(t)=1.8$, is introduced on day 30. 

The left panel in Figure \ref{fig:piece_wise_c_decreasing} gives a new threshold distancing cost function $\overline{c}$ when the distancing cost function satisfies $c(t)=1.8$ for $t \geq 30$ (the solid curve). After the introduction of the social-distancing measure, the prevalence decreases, and the new threshold $\overline{c}$ endogenously decreases as well. The figure shows that on day 50, the threshold cost is close to (but above) the current distancing cost. 

\begin{figure}[t]
\begin{minipage}{0.32 \hsize}
\begin{center}
\includegraphics[scale=0.38]{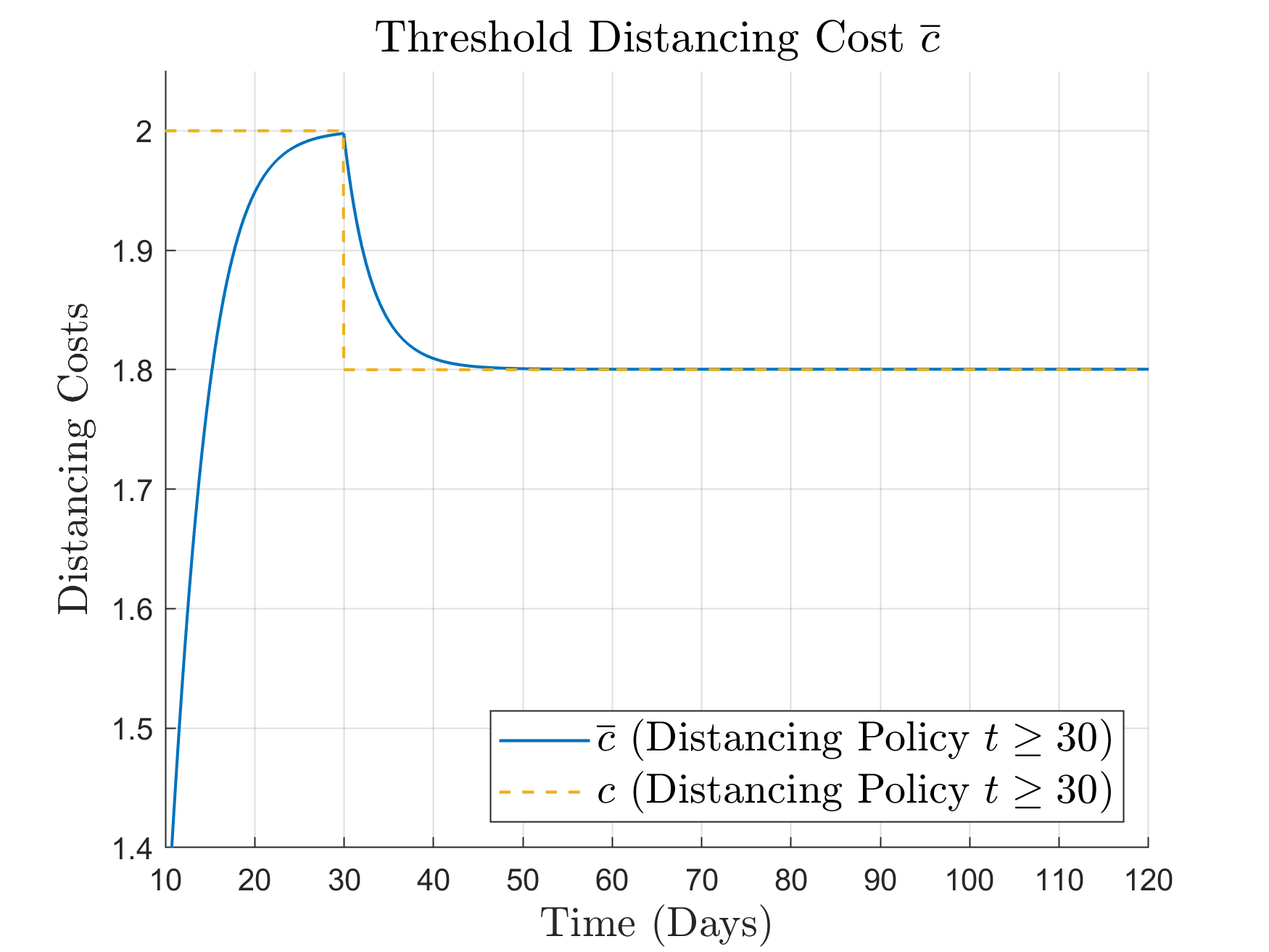}
\end{center}
\end{minipage}
\begin{minipage}{0.32 \hsize}
\begin{center}
\includegraphics[scale=0.38]{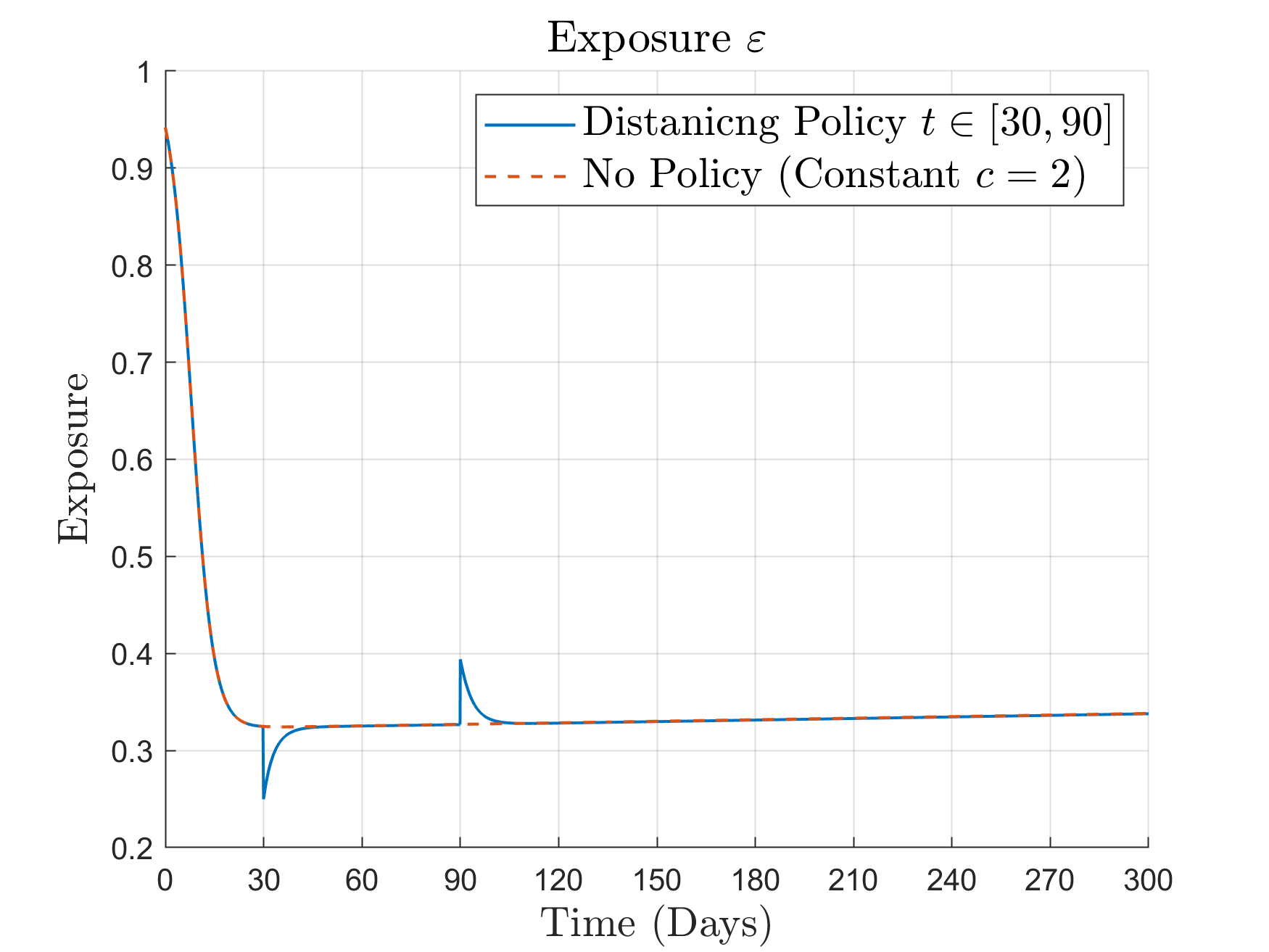}
\end{center}
\end{minipage}
\begin{minipage}{0.32 \hsize}
\begin{center}
\includegraphics[scale=0.38]{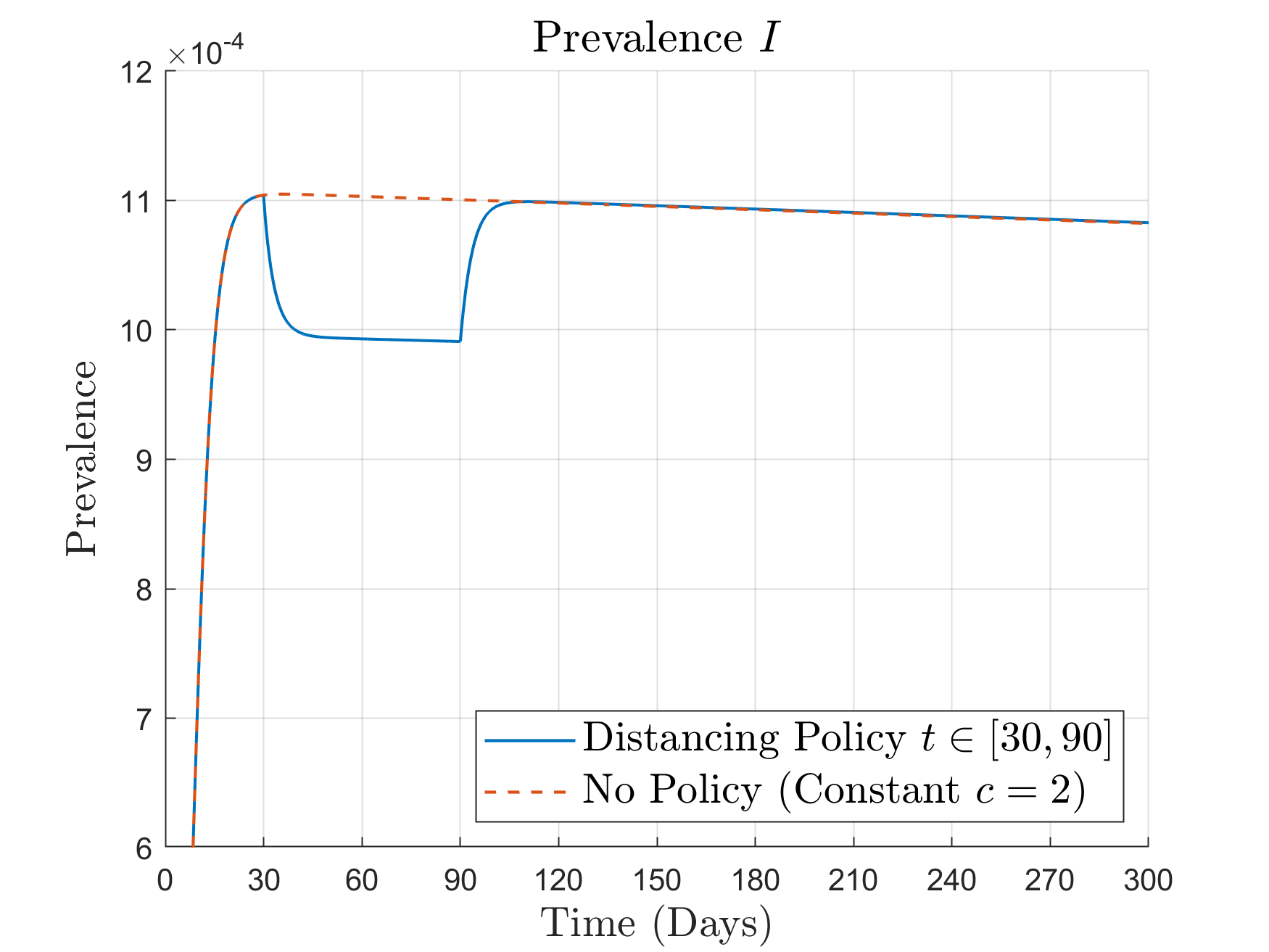}
\end{center}
\end{minipage}
\caption{\emph{Social-Distancing Policy}. The left panel depicts the threshold distancing cost function $\overline{c}$ over time. The central panel depicts the exposure level $\varepsilon$ over time. The right panel depicts the prevalence $I$ over time.}\label{fig:piece_wise_c_decreasing}
\end{figure}

To understand the new threshold distancing cost function, consider how individuals respond to the social-distancing measure. As the central panel of Figure \ref{fig:piece_wise_c_decreasing} shows, individuals best respond to the policy measure by decreasing their exposure levels. The resulting increase in distancing lowers prevalence, which leads to a feedback effect of increasing exposure. The prevalence nevertheless continues to decrease but individuals' responses slow down this decrease. The right panel illustrates this point. 

After day 50, virtually any easing of the social-distancing measure causes the second wave.\footnote{\citet{Anderson_20}, in discussing mitigation measures for suppressing the course of the COVID-19 pandemic, point out that interventions that reduce transmission greatly, which make the epidemic curve longer and flatter, cause the risk of a resurgence when
interventions are lifted. Our analysis sheds light on the following two additional insights. The first is that even a slight easing of the social-distancing measure may cause the second wave. The second is a behavioral mechanism behind the second wave through the threshold distancing cost function: how such measures also lower the threshold distancing cost to suppress the infection so as not to rise again.} For instance, assume that the distancing measure is lifted in its entirety after two months (i.e., on day 90). The right panel of Figure \ref{fig:piece_wise_c_decreasing} indicates that the infection resurges, and around day 111, the prevalence almost coincides with the case in which no distancing measure is introduced.\footnote{\label{fn:vaccine}A vaccination campaign (i.e., a reduction in $S$) during a lockdown helps prevent the resurgence of the infection. All else being equal, a reduction in $S(t)$ increases the threshold distancing cost $\overline{c}(t)$ as $\frac{\partial \overline{c}(t)}{\partial S(t)} = - \frac{\beta^{2}\eta \gamma I(t)}{(\beta S(t)-\gamma)^{2}}<0$.} 

\end{example}

\subsection{Public Policies as Time-Varying Distancing Cost}\label{sec:publicpoliciescost}

Several papers have analyzed optimal mitigation policies in a reduced form by assuming that a planner controls the path of the disease.\footnote{See, for example, \citet{acemoglu2020multi}, \citet*{alvarez2021simple}, \citet*{Farboodi_2021_internal}, and \citet{Kruse_Strack_22}.} That is, they assume that a planner directly controls the time-varying transmission rate, $\beta(t)$, or social interactions, $\varepsilon(t)$. Both cases can be viewed as a planner controlling an effective transmission rate $\tilde{\beta}(t)=\beta(t) \varepsilon(t)$, in which either $\beta(t)$ is controlled and $\varepsilon(t)$ is constant or in which $\beta$ is constant and $\varepsilon(t)$ is controlled.

While mask mandates can directly affect the transmission rate, the attainable levels of the transmission rate are limited by such a policy alone. Many countries have introduced additional policies beyond mask mandates to reduce transmission during the COVID-19 pandemic. Such policies aim at reducing the spread of the infection via reduced social interactions, such as bar and restaurant closures. However, to understand the effect of such policies---which affect the individuals' incentives to socially distance---, behavior should be modeled explicitly because they are only effective through individuals' endogenous choices, which, in turn, depend on the current state of the epidemic.\footnote{\citet*{Carnehl_Fukuda_Kos_23} show that policies that affect distancing incentives via reductions in the transmission rate and changes in the cost of distancing have qualitatively different effects on the path of an epidemic.} 

Nevertheless, the results obtained in these papers are important to understand the desirable epidemic paths of a planner who optimizes subject to macroeconomic or other cost considerations. Therefore, we show how our equilibrium distancing model with a time-varying cost can be used to back out the policy path affecting the cost directly to induce a desirable effective transmission rate $\tilde{\beta}(t)$. That is, given an optimal path $\tilde{\beta}(t)$, which was obtained without explicitly modeling behavior, we derive the cost function $c(t)$ that can implement the time-varying transmission path when endogenous behavioral responses are taken into account.

Consider a desirable time-varying transmission rate $\tilde{\beta}(t)$ for given primitives of the non-behavioral SIR model (i.e., $\beta$, $\gamma$, $I_{0}$, and $S_{0}=1-I_{0}$). The dynamics of the disease under the desirable transmission rate function $\tilde{\beta}$ follows the system of equations (\ref{eq:S_dot}), (\ref{eq:I_dot}), and (\ref{eq:R_dot}) where $\beta \varepsilon(t)$ is replaced by $\tilde{\beta}(t)$. Then, we can use our model to solve for the time-varying distancing cost function $\tilde{c}$ implementing  the transmission rate function $\tilde{\beta}(t) = \beta \varepsilon(t)$ via
\begin{align*}
\tilde{c}(t):=\frac{\beta^2\eta I(t)}{\beta-\tilde{\beta}(t)},
\end{align*}
provided $\tilde{\beta}(t) < \beta$. Without distancing fatigue, the required lockdown severeness $\tilde{\ell}(t)$ follows
\begin{align*}
\tilde{\ell}(t):=\frac{\beta^2\eta I(t)}{c_0(\beta-\tilde{\beta}(t))}.
\end{align*}

However, we can straightforwardly also incorporate distancing fatigue $\varphi(t)$ to obtain:
\begin{align*}
\tilde{\ell}_\varphi(t):=\frac{1}{c_0}\left(\frac{\beta^2\eta I(t)}{\beta-\tilde{\beta}(t)}-\varphi(t)\right).
\end{align*}

It should be noted that there is an endogenous upper bound on the implementable $\tilde{\beta}(t)$, which derives from individuals' endogenous distancing without policy interventions. Unless meetings can be subsidized during an epidemic---that is, more exposure encouraged than individuals would voluntarily engage in---, $\tilde{\beta}(t)>\beta$ cannot be attained. 

An important observation is that the strictness of the policies in place depends not only on the transmission rate to be implemented but also on current prevalence $I(t)$, fatigue $\varphi(t)$, and the cost of infection $\eta$. If the prevalence or the cost of infection is high or if fatigue is low, policies do not have to be as strict to induce a certain transmission rate as otherwise. This suggests that in models studying the optimal control of a transmission rate during an epidemic, the cost function of reducing the transmission rate should at least depend on the current prevalence to take endogenous distancing decisions into account.\footnote{The distancing cost function $\tilde{c}$ can be interpreted as the ratio between the marginal benefit of distancing $\beta I(t) \eta$ (relative to $c_0$) and the reduction of the transmission rate $\frac{\beta - \tilde{\beta}(t)}{\beta}$.}

Finally, an analogous approach is feasible to implement desired levels of the effective reproduction number $\mathcal{R}^{e}(t)$ which measures how many secondary infections are caused by each infected individual.\footnote{In the non-behavioral SIR model, the effective reproduction number is given by $\frac{\beta}{\gamma} S(t)$. In our behavioral SIR model, the effective reproduction number is given by $\frac{\beta \varepsilon(t)}{\gamma} S(t)$.} Whenever $\mathcal{R}^{e}(t)>(<)1$, the prevalence is increasing (decreasing). For example, \citet{Budish_2020} considers $\mathcal{R}^{e}(t) \leq 1$ as a constraint for a planner without an explicit dynamic equilibrium model.\footnote{Note that while the effective reproduction number empirically has the tendency to be close to one for some time endogenously (see, for example, \citet*{atkeson2020four} for data on the COVID-19 pandemic, and the discussion in \citet{Gans_22}) policymakers have to take into account how changes in the distancing cost due to imposing and lifting lockdowns affect the prevalence.} In our setting, this constraint corresponds to the current policy satisfying the constraint that $c(t)\leq \overline{c}(t)$ as in Proposition \ref{prop:threshold_cost}.

\section{Numerical Analysis of Public Policies with Distancing Fatigue}\label{sec:lockdown_and_fatigue}

In this section, we combine the two sources of time variation in the distancing cost\textemdash public policies and distancing fatigue\textemdash and demonstrate that distancing fatigue can have adverse effects on a well-intended public policy. 

We simulate our model on the basis of numbers motivated by China's strict COVID-19 lockdown and show that a strict lockdown from the outset of the epidemic may increase both the final number of infected individuals and the peak of the prevalence in the second wave arising upon the lifting of the lockdown. The reason is that a lockdown imposed at the beginning of the epidemic, that does not completely eradicate the infection, effectively postpones the spread of the disease until the lifting of the lockdown. At that point, individuals are fatigued and thus reluctant to distance as much as they would have in the absence of the lockdown.


While the analysis based on China's lockdown below is relatively extreme in terms of the strictness and duration of the policy as well as the calibrated fatigue parameters, the insights are more general. Appendix \ref{sec:moderate_lockdown} provides an illustration of the disease dynamics for more moderate fatigue parameters and shorter, less strict policies, and demonstrates that similar qualitative patterns arise.


\begin{example}\label{exl:c_lockdown}
To illustrate the interaction between fatigue and public policy, we approximate China's COVID-19 lockdown within our model. We choose the model parameters $(\beta, \gamma, I_{0}, \eta, c_{0})$ based on the parameters calibrated in \citet*{Carnehl_Fukuda_Kos_23}. We impose a lockdown in the model starting 10 days after the epidemic's start. To focus on the effect of distancing fatigue during the lockdown, we assume that individuals' distancing cost is constant before the lockdown is imposed and follows equation (\ref{eq:c_distancing}) afterwards. For simplicity, we assume that the lockdown lasts for 365 days. The lockdown induces a 75\% reduction in social activity in line with the empirical findings in \citet{zhong2022covid}, who find a 74.1-80\% reduction in mobility in China. 
 To obtain reasonable distancing fatigue parameters $(k,r)$ for equation (\ref{eq:c_distancing}), we choose the fatigue parameters such that the model-predicted peak after the lifting of the lockdown would match the peak of the second wave observed in China. We chose $k=0.02 \eta$ and $r = 0.01$. 

\begin{figure}
\begin{minipage}{0.32 \hsize}
\begin{center}
\includegraphics[scale=0.38]{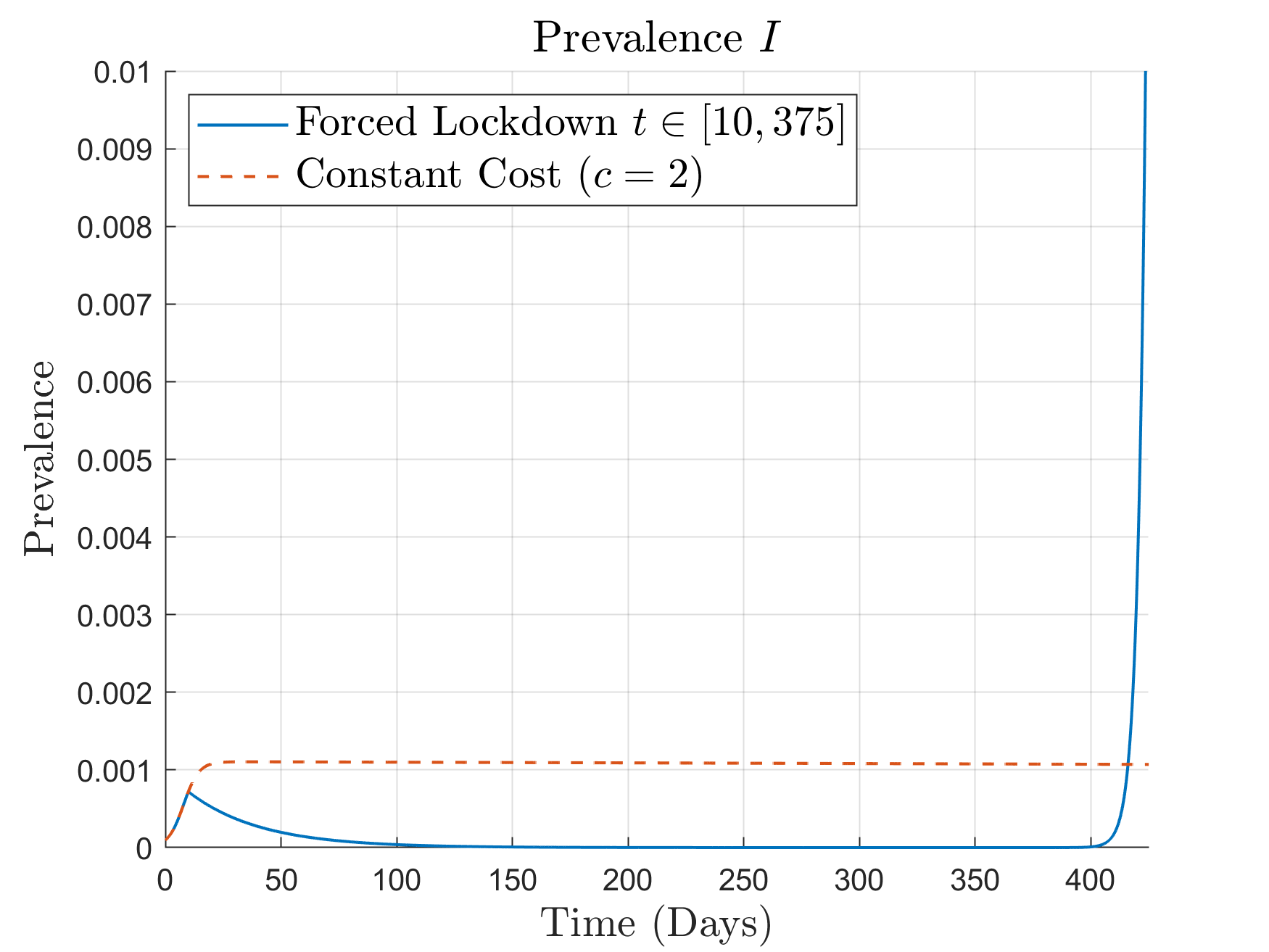}
\end{center}
\end{minipage}
\begin{minipage}{0.32 \hsize}
\begin{center}
\includegraphics[scale=0.38]{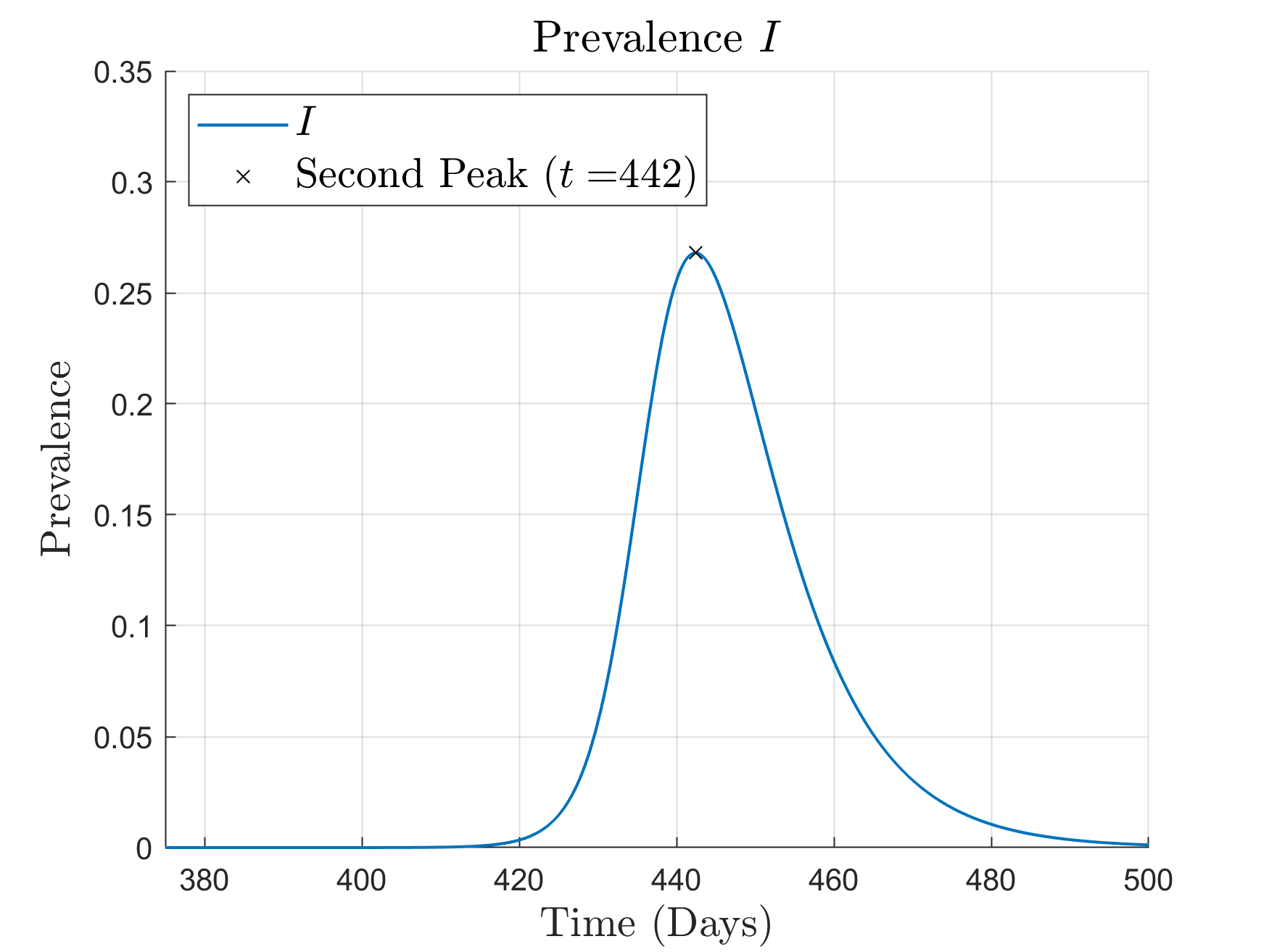}
\end{center}
\end{minipage}
\begin{minipage}{0.32 \hsize}
\begin{center}
\includegraphics[scale=0.38]{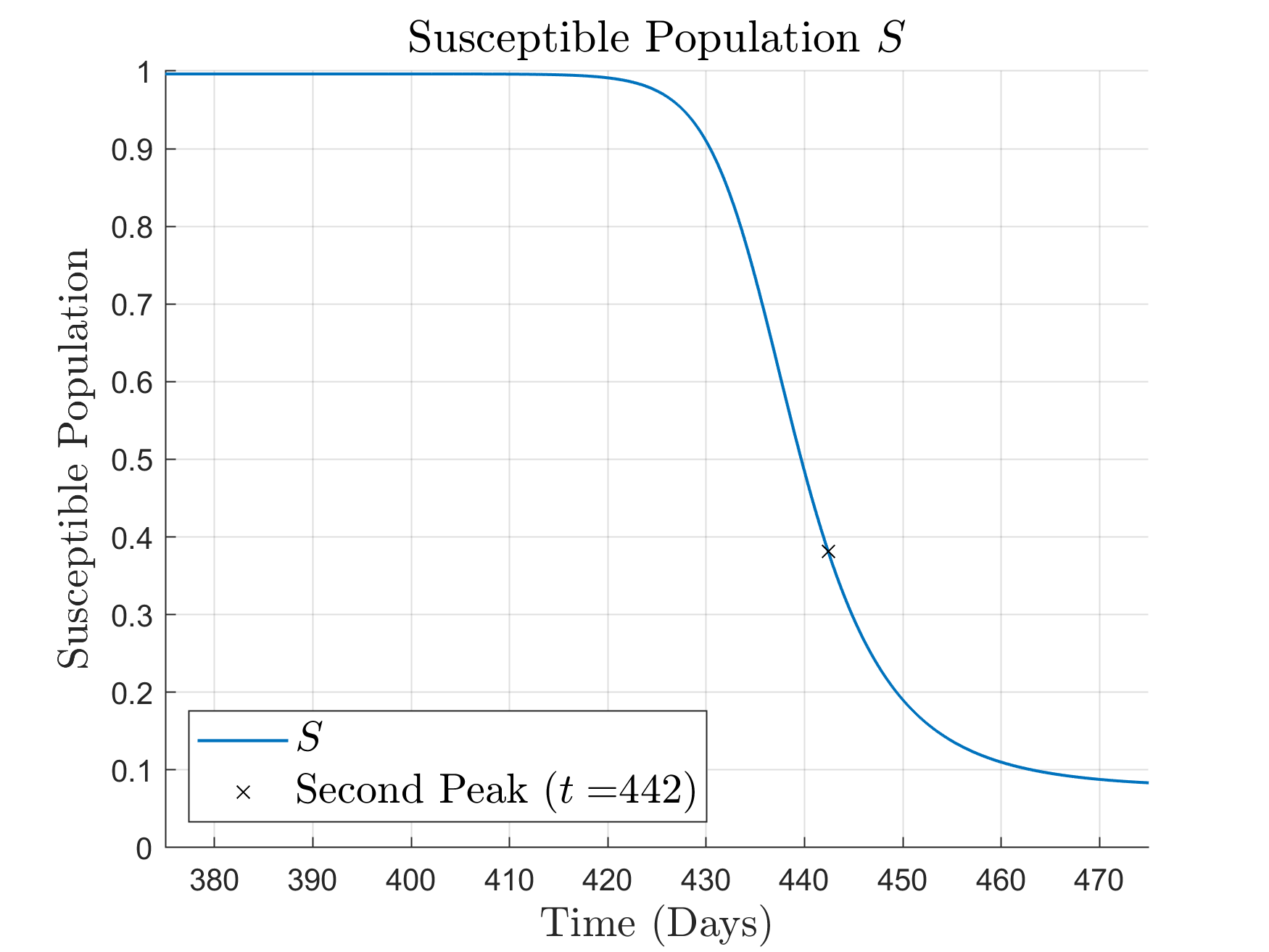}
\end{center}
\end{minipage}
\caption{\emph{Lockdown Policy}. The left panel depicts the prevalence curves (with and without the lockdown) until after 50 days at which the lockdown is lifted. The central panel depicts the prevalence $I$ since the lockdown is lifted. The right panel depicts the susceptible population $S$ since the lockdown is lifted.}\label{fig:c_lockdown}
\end{figure}

The left panel of Figure \ref{fig:c_lockdown} depicts the prevalence curves under no lockdown policy (dashed) and under the lockdown policy (solid) until day 425 (50 days after lifting the lockdown). The central panel of Figure \ref{fig:c_lockdown} depicts the prevalence curve after the lockdown policy is lifted (day 375). The prevalence peaks on day 442, approximately two months after the policy is lifted. At the peak, around 27\% of individuals are infected. The right panel suggests that, at that point, approximately two thirds of the population has been infected.\footnote{This is not far away from the statement made by the chief epidemiologist of China's Center for Disease Control and Prevention that the ``epidemic has already infected about 80\% of the people'' in China as of January 21, 2023 (https://edition.cnn.com/2023/01/22/china/china-covid-80-lunar-new-year-intl-hnk/index.html).}

To understand the role played by the strict lockdown interacting with distancing fatigue, note that the non-behavioral SIR model, which provides an upper bound for our behavioral model, predicts the peak prevalence at roughly 31\%.\footnote{A strict, long lockdown leads to high levels of fatigue and thus a high distancing cost upon lifting the lockdown. When $\beta\eta/c(t) \approx 0$, the disease dynamics after the lockdown can be well approximated by the standard non-behavioral SIR model with the post-lockdown initial condition $(S_\ast,I_\ast)$ as the share of susceptible individuals is still high as the lockdown was imposed early in the epidemic. This model predicts a peak prevalence of
\begin{equation*}\label{eq:second_peak_prevalence}
\overline{I} = \frac{\gamma}{\beta} \log \left( \frac{\gamma}{\beta} \right) - \frac{\gamma}{\beta} - \frac{\gamma}{\beta} \log (S_{\ast}) + S_{\ast} + I_{\ast}
\end{equation*}
(see, for instance, \citealp{brauer2012mathematical}). As discussed in the main text, this gives the prevalence of the second peak at $31\%$ as opposed to $27\%$ in our example.} In a behavioral SIR model without distancing fatigue, the peak prevalence would be at about $0.1\%$. Hence, it seems that in a model with both distancing fatigue and a long and strict lockdown the benefits of voluntary social distancing are almost entirely removed. 

We can quantify the contribution of the prolonged lockdown on the peak prevalence by considering the situation in which the lockdown is not implemented but individuals still accumulate fatigue. In this case, the peak prevalence would be at about $15\%$ (see the right panel of Figure \ref{fig:c_diff_lockdowns}). Thus, while fatigue alone accounts for an increase of the peak prevalence from $0.1\%$ to $15\%$, the prolonged lockdown accounts for an additional increase of the peak prevalence from $15\%$ to $27 \%$, suggesting that the contribution of the prolonged lockdown is substantial.


\begin{figure}
\begin{minipage}{0.49 \hsize}
\begin{center}
\includegraphics[scale=0.55]{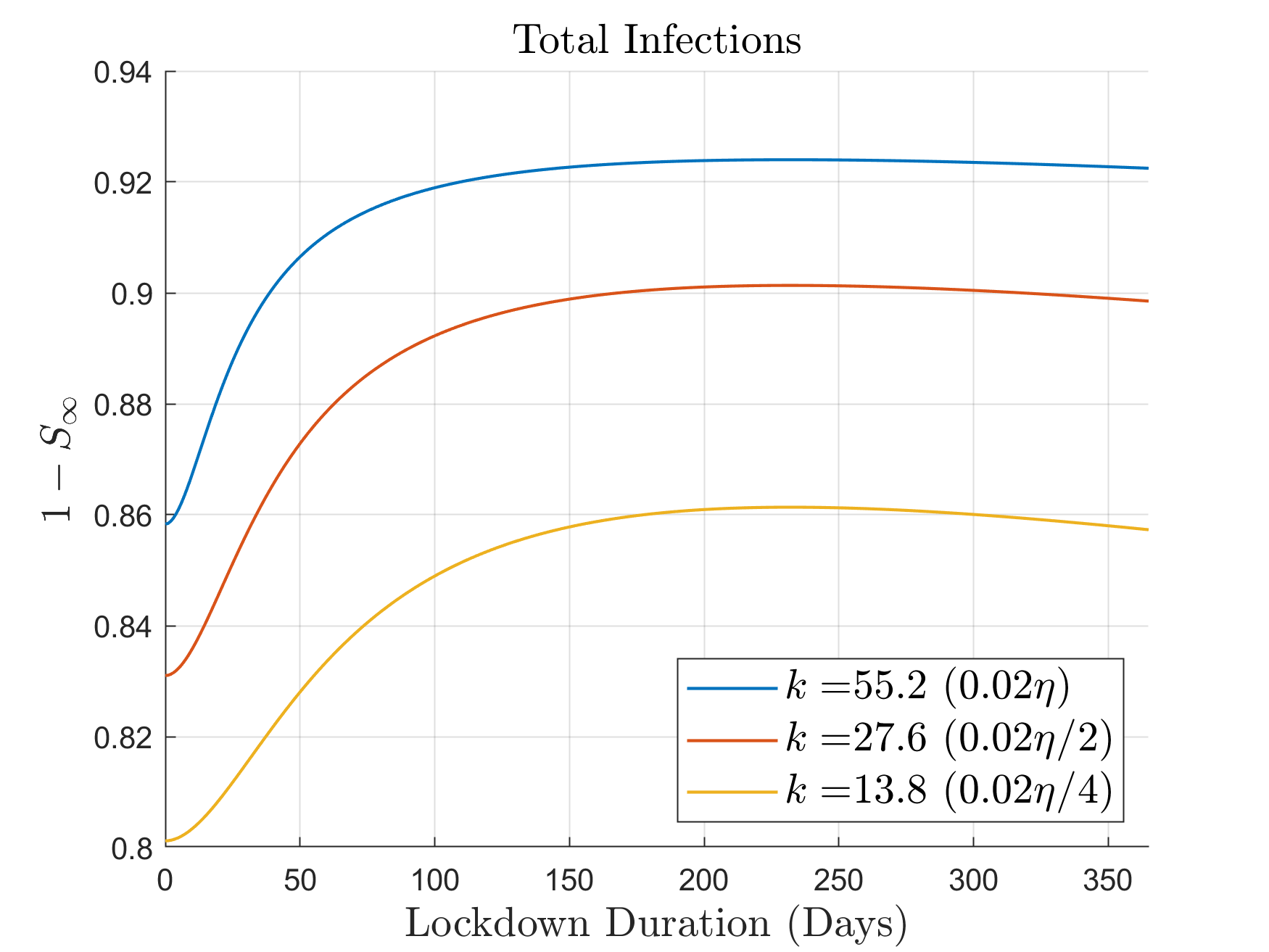}
\end{center}
\end{minipage}
\begin{minipage}{0.49 \hsize}
\begin{center}
\includegraphics[scale=0.55]{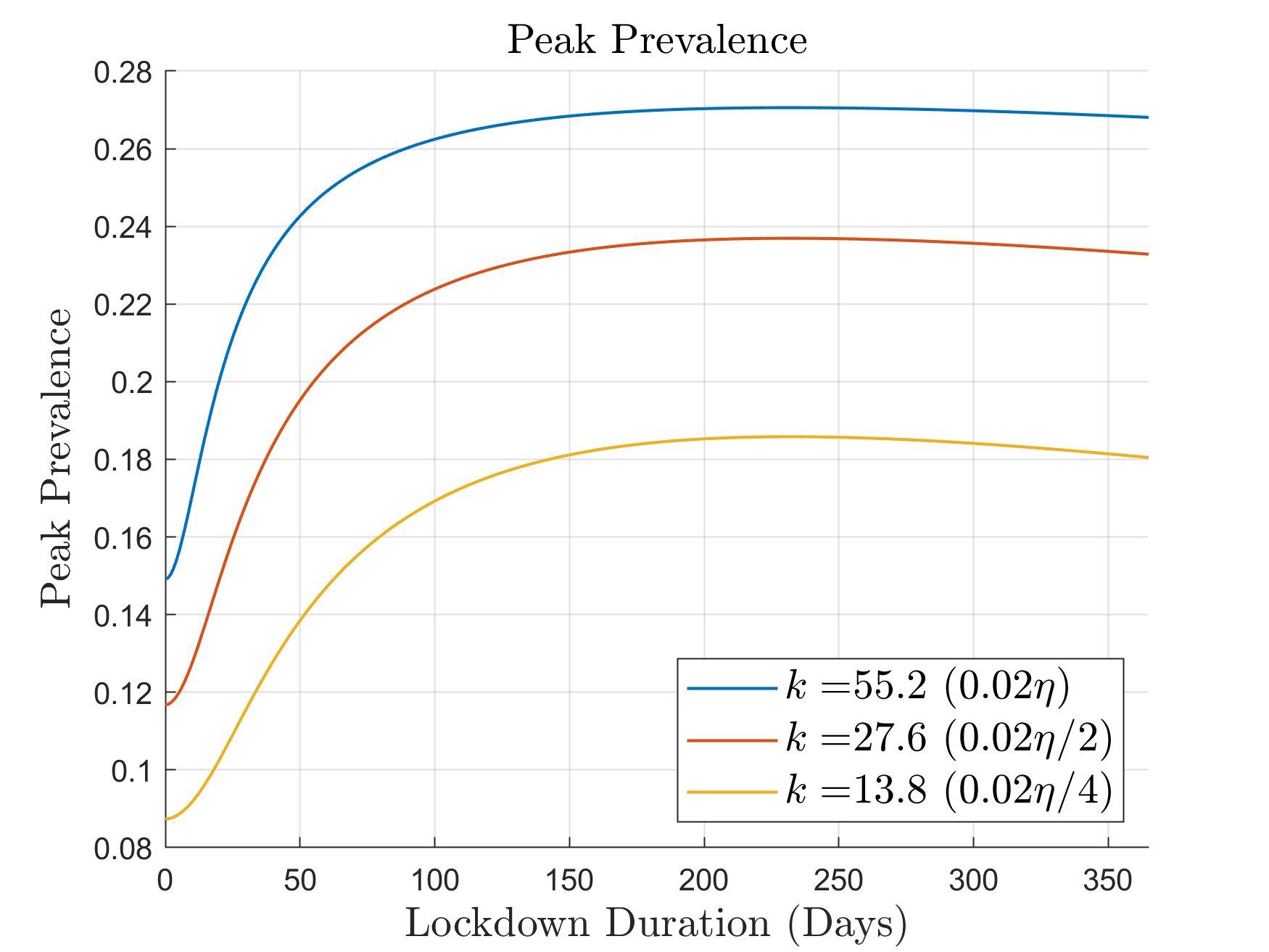}
\end{center}
\end{minipage}
\caption{\emph{Different Lockdown Policies}. The left panel depicts the total number of infections ($1-S_{\infty}$) as a function of the lockdown duration and for different starting times of the lockdown. The right panel depicts the corresponding levels of peak prevalence.}\label{fig:c_diff_lockdowns} 
\end{figure}

Figure \ref{fig:c_diff_lockdowns} depicts the total number of infections ($1-S_{\infty}$) and the prevalence at the second peak as a function of the lockdown duration. To see the robustness of the qualitative features of these two measures, we consider three different fatigue accumulation rates $k \in \{ 0.02 \eta, 0.01 \eta, 0.005 \eta \}$. The other model parameters are fixed. It is striking to see that both the total number of infections $1-S_{\infty}$ and the peak prevalence are lowest without any lockdown---they are initially increasing in the lockdown duration. After a certain threshold duration of the lockdown, we observe that while a policy-maker may have preferred to impose a shorter or no lockdown at all, the size of the epidemic and the second peak prevalence can be reduced only by keeping the lockdown in place for an extended period of time.\footnote{Note, however, that in these simulations, we do not impose any direct lockdown cost and only consider relatively early starting dates of the lockdown.}

This example illustrates the important trade-off between breaking an initial wave of an epidemic with a strict lockdown policy and the cost of lifting the lockdown while individuals have accumulated substantial levels of distancing fatigue.
\end{example}

\section{Conclusion}


This paper introduces a behavioral SIR model with time-varying distancing costs. After deriving qualitative results about the course of an epidemic with distancing cost functions that vary over time, we focus on two main applications: distancing fatigue and public policies. 

We incorporate endogenously evolving distancing fatigue into a behavioral SIR model by assuming that individuals' distancing costs increase in their past distancing behavior. We show that distancing fatigue alone cannot cause a second wave of infection. In fact, for a second wave to arise, the distancing cost has to increase rapidly after the first peak of active infections. Distancing fatigue postpones the time at which prevalence peaks and raises the level of peak prevalence. Thus, distancing fatigue may have substantial consequences for the medical system even though the qualitative patterns of the infection dynamics remain the same as in the model without fatigue (i.e., the prevalence remains single-peaked). 

While distancing fatigue alone does not cause a second wave, changes in public policies can. In particular, the removal of a mitigation policy can induce a sufficient increase in the distancing cost for a second wave to arise. Thus, policymakers must consider the consequences of changes in public policies through behavioral responses carefully. To guide such considerations, we formulate a threshold distancing cost function: If individuals' distancing costs remain below the threshold then the prevalence does not increase. 

Finally, we examine the interplay of lockdown policies and distancing fatigue. Crucially, distancing fatigue imposes a negative dynamic spillover on lockdowns. The policy that curtails mobility in the current period reduces distancing incentives in the future via two channels: i) lower prevalence and ii) accumulated distancing fatigue. Holding lockdown stringency fixed, distancing fatigue reduces lockdown effectiveness over time, and increases the prevalence level of a second wave should the lockdown be lifted. Consequently, a current lockdown decreases the effectiveness of any future lockdown policies. In addition, we demonstrate that longer lockdowns can cause higher prevalence levels in the second wave---even exceeding the prevalence levels without any lockdown at all.

\bibliographystyle{ecta}
\bibliography{covid_ref_2}

\newpage

\appendix

\section{Proofs}\label{sec:proofs}

\begin{proof}[\bf{Proof of Proposition \ref{prop:existence}}]
At each time $t$, an individual's problem (\ref{eq:naive_prob}) is concave. Thus, the first-order condition (\ref{eq:naive_distancing}) is sufficient. This pins down the individual's optimal distancing in the SIR dynamics. 

Using the exposure obtained from  (\ref{eq:naive_distancing}) in the SIR dynamics together with the cost-function evolution yields
\begin{align}
\dot{S}(t) & = - \beta S(t) I(t) \max \left( 1 - \frac{\eta \beta I(t)}{c(t)},0 \right), \label{eq:S_dot_e} \\
\dot{I}(t) & = \beta S(t) I(t) \max \left( 1 - \frac{\eta \beta I(t)}{c(t)},0 \right) - \gamma I(t), \label{eq:I_dot_e} \\
\dot{R}(t) & = \gamma I(t), \label{eq:R_dot_e} \\
\dot{c}(t) & = F \left( t, c(t), \max \left( 1 - \frac{\eta \beta I(t)}{c(t)},0 \right) \right), \label{eq:c_dot_e}
\end{align}
for all but possibly a finite number of $t$, at which at least one of the variables $(S,I,R,c)$ is not differentiable. Let $t_1 < \cdots < t_N$ be the set of these points (this set may possibly be empty). Let $t_{N+1} = \infty$.

Thus, in any equilibrium, $(S,I,R,c)$ is characterized by the system of differential equations $\frac{d}{dt}(S,I,R,c) = G(t,S,I,R,c)$, where $G$ is defined by (\ref{eq:S_dot_e}), (\ref{eq:I_dot_e}), (\ref{eq:R_dot_e}), and (\ref{eq:c_dot_e}). The initial condition is $(S(0),I(0),R(0),c(0)) = (S_{0}, I_{0},0,c_{0})$. Then, the initial value problem admits a unique solution $(S,I,R,c)$ on $[0, t_1)$, as the system satisfies the conditions of the Picard-Lindel\"{o}f Theorem. Namely, the function $G$ is continuous on the domain $D=[0, t_1) \times [0,1]^{3} \times [\underline c, \infty)$, and $G$ is uniformly Lipschitz continuous in $(S,I,R,c)$: there exists a Lipschitz constant $L$ satisfying $\| G(t,S,I,R,c) - G(t,\tilde{S}, \tilde{I}, \tilde{R}, \tilde{c}) \| \leq L \| (S, I,R,c) - (\tilde{S}, \tilde{I}, \tilde{R}, \tilde{c}) \|$ for each $t \in [0, t_1)$. See, for example, \citet*{Walter_98}. Since the equilibrium definition requires $S$, $I$ and $R$ to be continuous, we apply the same logic to the interval $[t_1,t_2)$ with the initial value $(S(t_1),I(t_1),R(t_1))= \displaystyle \lim_{t \uparrow t_1} (S(t),I(t),R(t))$ and all the subsequent intervals. Note that each $c(t_{n})$ is also given. Now, $\varepsilon = \varepsilon_{i}$ is uniquely determined, and hence the model admits a unique and symmetric equilibrium.

Next, we show $\displaystyle \lim_{t \rightarrow \infty} I(t) = 0$. Since $R (\cdot) \in [0,1] $ is weakly increasing, $\displaystyle \lim_{t \rightarrow \infty} R(\infty)$ exists in $[0,1]$. By equation (\ref{eq:R_dot_e}), we must have $0 = \displaystyle \lim_{t \rightarrow \infty} \dot{R}(t) = \gamma \lim_{t \rightarrow \infty} I(t)$, establishing $I_{\infty} =0$. 

Next, $\displaystyle \lim_{t \rightarrow \infty} \varepsilon(t) =1$ follows from taking the limit of (\ref{eq:naive_distancing_symmetric}) because the distancing cost is bounded from below, $c(t)\geq \underline{c}$, and $\displaystyle \lim_{t \rightarrow \infty} I(t) = 0$. 

Finally, having established $\displaystyle \lim_{t \rightarrow \infty} I(t) = 0$ and $\displaystyle \lim_{t \rightarrow \infty} \varepsilon(t) =1$, one can show $S_{\infty} \in \left( 0, \frac{\gamma}{\beta} \right)$ as in the proof of Lemma 3 in \citet*{Carnehl_Fukuda_Kos_23}. 
\end{proof}

\begin{proof}[\bf{Proof of Proposition \ref{prop:condition_one_wave}}]
To have at least two peaks, there must be two local strict maxima of $I$ at $t_{2} > t_{1} \geq 0$. Because $I$ is continuous it has a minimum on $[t_1,t_2]$ by the extreme value theorem. Moreover, since $t_{1}$ and $t_{2}$ are local strict maxima, the minimum has to be attained at some $\hat t \in (t_{1},t_{2})$. The fact that in equilibrium $S,~I$ and $R$ are continuous implies that $I$ is differentiable and that its derivative is given by (\ref{eq:I_dot}).  

As $I$ has a local minimum at $\hat t$, $\dot{I}(\hat t)=0$ and thus $\beta \varepsilon(\hat t) S(\hat t) = \gamma$. It follows that $\varepsilon (t) = 1- \frac{\beta \eta I(t)}{c(t)}>0$ and $S(t)>0$ in the neighborhood of $\hat t$. Therefore, the function $\ddot{I}$ exists and is obtained by differentiating $\dot{I}$ at $t = \hat{t}$:
\begin{equation*}
\ddot{I}(t) = \dot{I}(t)(\beta \varepsilon(t) S(t)-\gamma)+\beta I(t)(\dot{\varepsilon}(t) S(t)+\varepsilon(t)\dot{S}(t)).
\end{equation*}
Evaluating $\ddot{I}(t)$ with $\dot I(t) =0$ yields
\begin{equation}\label{eq:ddotI0}
\left.\ddot{I}(t)\right|_{\dot{I}(t)=0} = \beta I(t)(\dot{\varepsilon}(t) S(t)+\varepsilon(t)\dot{S}(t)).
\end{equation}
Substituting
\begin{equation*}
\dot{\varepsilon}(t)=\frac{- \beta \eta I(t)}{c(t)}\left(\frac{\dot{I}(t)}{I(t)}-\frac{\dot{c}(t)}{c(t)} \right) \text{ and } \dot{S}(t)=-\beta \varepsilon(t) I(t)S(t)
\end{equation*}
into (\ref{eq:ddotI0}) results in
\begin{equation}\label{eq:ddotI0_rewrite}
\left.\ddot{I}(t)\right|_{\dot{I}(t)=0} = \beta^2 \eta I^{2}(t)S(t) \left( \frac{\dot{c}(t)}{c^{2}(t)} - \frac{\varepsilon^{2}(t)}{\eta} \right).
\end{equation}

For no interior minimum to exist it is sufficient to show that $\ddot{I}(t)|_{\dot{I}(t)=0}<0$ for all $t >0$, which occurs precisely when (\ref{eq:condition_one_wave}) holds.  

Since $\varepsilon(t)=\frac{\gamma}{\beta S(t)}$ when $\dot{I}(t)=0$, (\ref{eq:condition_one_wave}) can be rewritten as
\begin{align*}
\frac{\dot{c}(t)}{c^{2}(t)} < \frac{\gamma^{2}}{\eta \beta^{2} S^{2}(t)} \text{ for all } t>0.
\end{align*} 
Due to $S$ being non-increasing over time, a sufficient condition for the above condition is (\ref{eq:condition_one_wave_2}), as desired. 
\end{proof}

\begin{proof}[\bf{Proof of Lemma \ref{lmm:continuous_c_sufficient}}]
Suppose $\sigma(t) > \frac{1}{\eta}c(t)$ for all $t>0$, that is, 
\begin{equation*}
\frac{1}{\eta} < \frac{\sigma(t)}{c(t)} = - \frac{d}{dt} \left( \frac{1}{c(t)} \right) \text{ for all } t>0.
\end{equation*}
Since $c$ is continuously differentiable, the right-hand side of the above expression is continuous. Integrating both sides from some $t_{0} >0$ to $t >t_{0}$, it follows from the fundamental theorem of calculus that 
\begin{equation*}
\frac{t-t_{0}}{\eta} + \frac{1}{c(t)} < \frac{1}{c(t_{0})}. 
\end{equation*}
Since $\frac{1}{c(t_{0})}$ is finite, the inequality can obtain only if $c(t) < 0$ from some $\hat{t}$ on. 
\end{proof}

\begin{proof}[\bf{Proof of Lemma \ref{lm:varepsilon_positive}}]
Suppose, to the contrary, that individuals choose $\varepsilon(t)=0$ for some $t >t'$, and let $\underline{t}:= \inf \{ t \geq t' \mid \varepsilon (t) =0\}$. Since $I$ and $c$ are continuous in equilibrium, so is $\varepsilon$. Thus, $\varepsilon(\underline{t})=0$ and consequently $\underline{t} > t'$. Towards the contradiction we will argue that in any small enough left neighborhood of $\underline{t}$, $\dot{\varepsilon}(t) >0$. 

At any $t$ where $\varepsilon (t) >0$, $\varepsilon$ is differentiable with derivative
\begin{align}\label{eq:dot_varepsilon}
\dot \varepsilon(t) = (1-\varepsilon(t))\left(\frac{\dot c(t)}{c(t)} - \frac{\dot I(t)}{I(t)}\right).
\end{align}
In addition, $\varepsilon(t) >0$ on $(\underline{t}, t')$ implies
\begin{align*}
c(t)-c_{0} &= k \int_{0}^{t} e^{-r(t-\tau)} (1-\varepsilon(\tau))d\tau\\
&<k \int_{0}^{t} e^{-r(t-\tau)} d\tau\\
&<\frac{k}{r}.
\end{align*}
Since $I$ and $c$ are continuous in equilibrium, for any $\delta_1>0$, there exists $\delta_2 >0$ such that $\varepsilon(t) < \delta_1$ if $t \in (\underline t - \delta_2, \underline t]$. But then, given that $c(t) - c_{0}< \frac{k}{r}$, $\delta_1$ can be chosen small enough so that $r(c(t)-c_{0}) < k(1-\varepsilon(t))$. In other words, for $\delta_2$ small enough, $\dot{c}(t) >0$ for $t \in (\underline t -\delta_2, \underline t)$. Moreover, equation (\ref{eq:I_dot}) implies that $\dot{I} <0$ whenever $\varepsilon < \frac{\gamma}{\beta}$.  Therefore, $\delta_1$ can be chosen so that $\dot{c}(t)>0$ and $\dot{I}(t) <0$. Consequently, due to (\ref{eq:dot_varepsilon}), $\dot{\varepsilon} (t) >0$ on $(\underline t - \delta_1, \underline t)$. But this means that, whenever $\varepsilon(t)$ becomes very small, it starts increasing and thus cannot reach $0$.
\end{proof}

\begin{proof}[\bf{Proof of Lemma \ref{lm:cI}}]
As $\dot{I}(0)>0$, it must be the case that  $\varepsilon (0) >0$. By Lemma \ref{lm:varepsilon_positive}, $\varepsilon(t) > 0$ for all $t \geq 0$. By implication $\varepsilon$ and therefore $\dot{c}$ are differentiable for all $t >0$.

Suppose $c$ attains a critical point at some $t$. Thus,  $\dot{c}(t)=0$.  Differentiating (\ref{eq:c_dot_discounting}) and evaluating it at $\dot{c}(t) =0$ yields
\begin{align*}
\left.\ddot{c}(t)\right|_{\dot{c}(t)=0}  = -k \dot{\varepsilon}(t) = k \frac{\beta \eta \dot{I}(t)}{c(t)}.
\end{align*}
Thus, if $c$ attains a local maximum (minimum) at $t$, it is necessary that $\dot{I}(t) \leq 0$ ($\geq 0$).
\end{proof}

\begin{proof}[\bf{Proof of Proposition \ref{prop:single_peak_discounting}}]
We first prove the first part of Proposition \ref{prop:single_peak_discounting}. Since $\dot{I}(0)>0$ and $I_{\infty} = 0$, it follows that $I$ peaks at least once. In addition, $\dot{I}(0)>0$ implies $\varepsilon (0) >0$ and thus by Lemma \ref{lm:varepsilon_positive}, $\varepsilon(t) > 0$ for all $t >0$. As a consequence, $\dot I(\cdot)$ is differentiable.

Let $t_{1}$ be some $t$ at which a local maximum of $I$ is attained. Thus, $\dot{I}(t_1)=0$ and $\ddot{I}(t_1) \leq 0$. Then by (\ref{eq:ddotI0_rewrite}) in the proof of Proposition \ref{prop:condition_one_wave}, it must be the case that
\begin{equation*}
\frac{\dot c(t_1)}{c^{2}(t_1)} \leq \frac{\varepsilon^{2}(t_1)}{\eta}.
\end{equation*}
Let $t_2$ be the smallest $t > t_1$ such that $\dot{I}(t) = 0$ and $\ddot{I}(t) \geq 0$. If it exists, $t_2$ is the first local minimum after $t_1$. If there is no local minimum after the first local maximum, our result is proven. 

We consider two cases. First, suppose $c(t_2) \geq c(t_1)$. Then:
\begin{align*}
\dot{c}(t_2) &= k(1-\varepsilon(t_2)) - r(c(t_2) -c(0)) \\
& < k(1- \varepsilon (t_1)) - r(c(t_1) - c(0))\\
&=\dot{c}(t_1),
\end{align*}
where the inequality follows from the fact that at any $t$ such that $\dot I(t) =0$, $\varepsilon(t)= \gamma/(\beta S(t))$ and that $S(t)$ is decreasing. As a consequence, 
\begin{equation*}
\frac{\dot c(t_2)}{c^{2}(t_2)} < \frac{\dot c(t_1)}{c^{2}(t_1)} \leq \frac{\varepsilon^{2}(t_1)}{\eta} < \frac{\varepsilon^{2}(t_2)}{\eta},
\end{equation*}
which, due to equality (\ref{eq:ddotI0_rewrite}) in the proof of Proposition \ref{prop:condition_one_wave}, contradicts the assumption that $\dot{I} (t_2) =0$ and $\ddot{I}(t_2) \geq 0$. 

Second, suppose $c(t_{2}) < c(t_{1})$. By the definition of $t_{2}$, $I(t_1) > I(t_2)$ and $I$ is decreasing on $[t_1,t_2]$. Since $c$ is continuous on $[t_{1},t_{2}]$, it attains a maximum and minimum on the interval by the extreme value theorem. Lemma \ref{lm:cI} implies that if $c$ attains an interior extremum, then it has to be a local maximum. Alternatively, $c$ is decreasing on the whole interval. In either case $\dot{c}(t_{2}) \leq 0$. But then the inequality $\frac{\dot{c}(t_2)}{c^2(t_2)} < \frac{\varepsilon^2(t_2)}{\eta}$ is automatically satisfied and thus $\ddot{I}(t_2) < 0$, which contradicts the supposition. 

Thus, there does not exist a time $t_{2} \in (t_{1}, \infty)$ such that $\dot{I}(t_{2}) =0$ and $\ddot{I}(t_{2}) \geq 0$. 

The third part of Proposition \ref{prop:single_peak_discounting} follows from Lemma \ref{lm:cI}. 

The proof of the second part of Proposition \ref{prop:single_peak_discounting} contains two steps.
The first step shows that $c$ has a local maximum. The second step shows that $c$ is single-peaked.

First, since $\dot{c}(0) = k (1-\varepsilon(0)) >0$, there exists $t_{1}$ such that $c(t_{1}) > c_{0}$. Otherwise, $\dot{c}(0) \leq 0$, a contradiction. Since $r>0$ and $\displaystyle \lim_{t \rightarrow \infty} \varepsilon(t)=1$, it follows that $\displaystyle \lim_{t \rightarrow \infty} c(t) = c_{0}$. Thus, there exists $t_{2} \geq t_{1}$ such that $c(t) \leq c(t_{1})$ for all $t \geq t_{2}$. Now, $c$ admits a local maximum on $[0, t_{2}]$ by the extreme value theorem. By construction, the local maximum of $c$ on $[0, t_{2}]$ is a local maximum on $[0, \infty)$. 

Second, suppose to the contrary that there would exist $t_1$ and $t_2$ at which $c$ attains local maxima and a $t' \in (t_1,t_2)$ such that $c(t') < \min \left( c(t_1),c(t_2) \right)$. Since $c$ is continuous, it has a minimum on the interval $[t_1,t_2]$ by the extreme value theorem. Let $\tilde t \in (t_1,t_2)$ be some $t$ at which $c$ is minimized over $[t_1,t_2]$. Then $\dot c(t_1) = \dot c(\tilde t) = \dot c(t_2)=0$. Using equation (\ref{eq:c_dot_discounting}), we obtain
\begin{equation*}
 1-\varepsilon(t) = \frac{r}{k}(c(t)-c_{0}) \text{ for } t \in \{t_1,\tilde t, t_2 \}.
\end{equation*}
As $r>0$, the inequality $c(\tilde t) < \min \left(c(t_1),c(t_2)\right)$ implies that $1-\varepsilon(\tilde t) < \min \left( 1-\varepsilon(t_1),1-\varepsilon(t_2) \right)$. In turn, the last two inequalities together with (\ref{eq:naive_distancing_symmetric}) imply that $I(\tilde t) < \min \left( I(t_1),I(t_2) \right)$, which would contradict that $I$ is single-peaked as established in Proposition \ref{prop:single_peak_discounting}.
\end{proof}

\begin{proof}[\bf{Proof of Proposition \ref{prop:threshold_cost}}]
Take $c$, $t$, and $c_{2}$ as in the statement of the proposition. Suppose that $c_2(t) = \overline{c}(t)$. Let $(S_2,I_2,R_2,c_2,\varepsilon_2)$ be the equilibrium under $c_2$. By the definition of the equilibrium, $S_2$ and $I_2$ are continuous. Notice that $S_2$ and $I_2$ coincide with $S$ and $I$ on $[0, t]$. 

If $c_2$ has any discontinuities at some $\tau > t$, let $t'$ be smallest $\tau > t$ where $c_2$ is discontinuous; otherwise set $t' = \infty$. Since $S_2$ and $I_2$ are continuous, it follows from (\ref{eq:S_dot}) and (\ref{eq:I_dot}) that $\dot{S}_2$ and $\dot{I}_2$ exist and are continuous on $(t, t')$. Notice that 
\begin{align*}
\beta \varepsilon_2 (t) S_2(t) - \gamma = 0.
\end{align*}
By continuity of $\varepsilon_2$ and $S_2$ for every $\delta_1>0$ there exists a $\delta_2 >0$ such that $|\varepsilon_2 (t) S_2(t) - \varepsilon_2 (s) S_2(s)| < \delta_1$ for all $s$ such that $|s - t| < \delta_2$. Consequently, 
\begin{align*}
\dot I_2(\tau)& = I_2(\tau) (\beta \varepsilon_2 (\tau) S_2(\tau) - \gamma)\\
&<I_2(\tau) (\beta \varepsilon_2 (t) S_2(t) + \delta_1- \gamma)\\
&=\delta_1 I_2(\tau),
\end{align*}
for all $\tau \in (t, t+ \delta_2)$. Therefore the right limit of $\dot I_2(\tau)$ at $t$ is 0. It is then easy to see that if $c_2(t) < \overline{c}(t)$ the derivative of $I_2$ would be smaller than $0$.
\end{proof}

\section{Appendix to Section \ref{sec:continuous_costs}: Applying Proposition \ref{prop:condition_one_wave}}\label{sec:dec_at_0}

We apply Proposition \ref{prop:condition_one_wave} to show that the prevalence of a disease is single-peaked for several natural distancing-cost function specifications. The examples illustrate the usefulness of Proposition \ref{prop:condition_one_wave} as it does not require an analysis of the potentially complicated disease dynamics but only an inspection of the cost function itself.


\begin{example}\label{exl:linear}
Suppose that the cost function is linear in time:
\begin{align*}
c(t) = c_{0} + k\cdot t, 
\end{align*}
where $c_{0}>0$ and $k>0$. In this case, the term $\frac{\dot{c}(t)}{c^{2}(t)} =\frac{k}{c^{2}(t)}$ is non-increasing over time. Therefore, as in the arguments in the proof of Proposition \ref{prop:condition_one_wave}, there do not exist two points in time $t_{1}$ and $t_{2}$ such that $t_{1} < t_{2}$, $\dot{I}(t_{1})= \dot{I}(t_{2}) =0$, and $\ddot{I}(t_{1})<0$ and $\ddot{I}(t_{2})<0$. Consequently, if $\dot{I}(0) >0$, prevalence is single-peaked.\footnote{More generally, the same argument holds for a distancing cost function that is concave in time.} 
\end{example}

\begin{example}\label{exl:simplified_fatigue}
Suppose that the cost function features distancing fatigue of the following form: 
\begin{align*}
	\dot{c}(t) = k (1-\varepsilon(t)) c(t) \text{ with } c(0) = c_0,
\end{align*}
where $k> 0 $ is a constant. In this formulation, individuals' distancing cost is increasing in the past distancing choices and non-decreasing over time. The sufficient condition implies that 
\begin{align*}
k \leq \frac{\varepsilon^2(t)}{1-\varepsilon(t)} \frac{c(t)}{\eta}.
\end{align*}
Note that the right-hand side can be bounded from below by $\underline{\varepsilon}^2 \frac{c_0}{\eta}$ where $\underline{\varepsilon}:= \min_t \varepsilon(t)$. Hence, whenever individuals never fully distance themselves, that is, whenever $\underline{\varepsilon}>0$, no second wave can arise if the accumulation rate $k$ is sufficiently low. 
\end{example}

\begin{figure}
\begin{minipage}{0.49 \hsize}
\begin{center}
\includegraphics[scale=0.55]{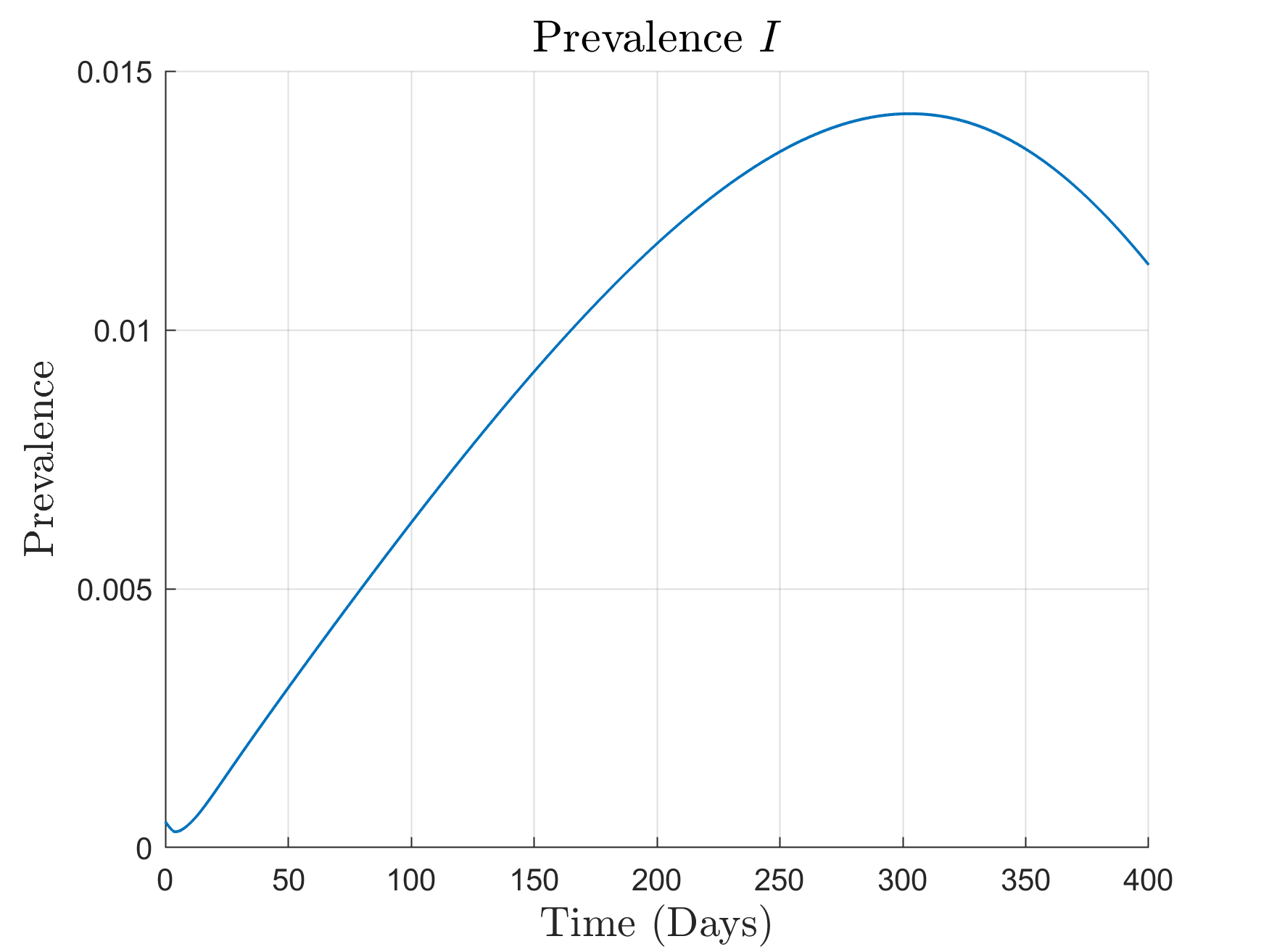}
\end{center}
\end{minipage}
\begin{minipage}{0.49 \hsize}
\begin{center}
\includegraphics[scale=0.55]{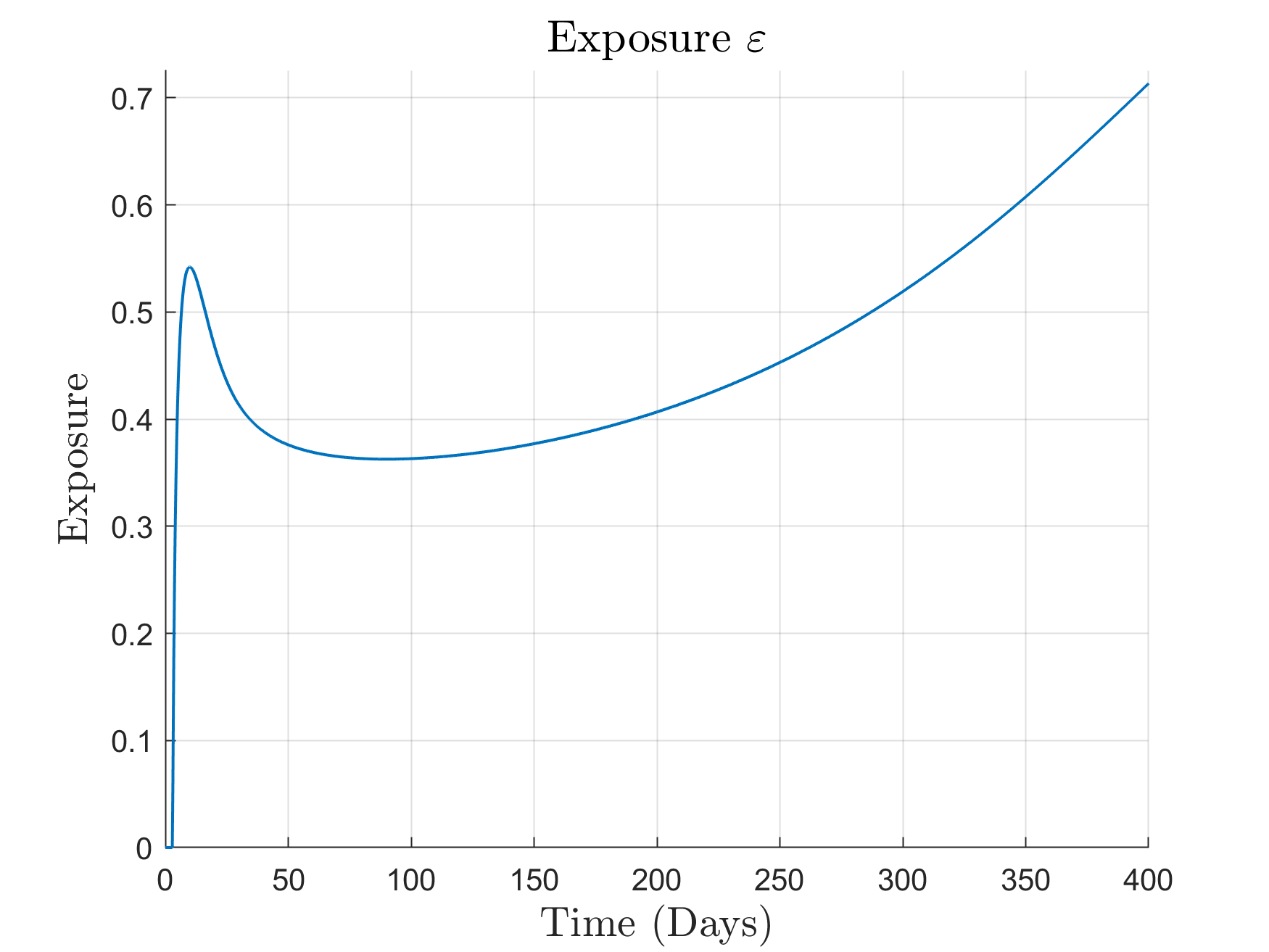}
\end{center}
\end{minipage}
\caption{\emph{Example \ref{exl:linear}}. The left panel depicts the prevalence $I$ over time. While $I$ initially decreases, it starts to increase at around $t=5$. The right panel depicts the exposure level $\varepsilon$ over time.} \label{fig:linear_exl}
\end{figure}

Note that an important qualification for the single-peakedness in the previous two examples is that $\dot{I}(0)>0$. If, however, the prevalence decreases at the outset, $\dot{I}(0)<0$, a minimum can arise when the distancing costs vary over time.\footnote{This is in contrast to the standard non-behavioral SIR model or the behavioral SIR model with constant distancing cost.} The right panel of Figure \ref{fig:linear_exl} illustrates this point in the context of Example \ref{exl:linear}.\footnote{The parameters are $(\beta, \gamma, \eta) = (0.3+\frac{1}{7}, \frac{1}{7}, 2761.63)$ and $(I_{0}, c_{0}, k) = (5 \times 10^{-4}, 0.05, 0.005)$. As discussed in footnote \ref{fn:calibration}, the parameters $(\beta, \gamma, \eta)$ are calibrated for the onset of COVID-19 (when the cost of distancing is normalized at $c=2$) as in \citet*{Carnehl_Fukuda_Kos_23}.} The prevalence decreases at the outset because it is not costly at all for the individuals to engage in distancing. However, since the prevalence decreases and the distancing costs increase, the individuals start increasing their exposure, which leads to an increase in the prevalence. 



\section{Appendix to Section \ref{sec:discountinuous_costs}}

\subsection{Effect of Holidays}\label{sec:holidays}

\begin{figure}[h]
\begin{minipage}{0.49 \hsize}
\begin{center}
\includegraphics[scale=0.55]{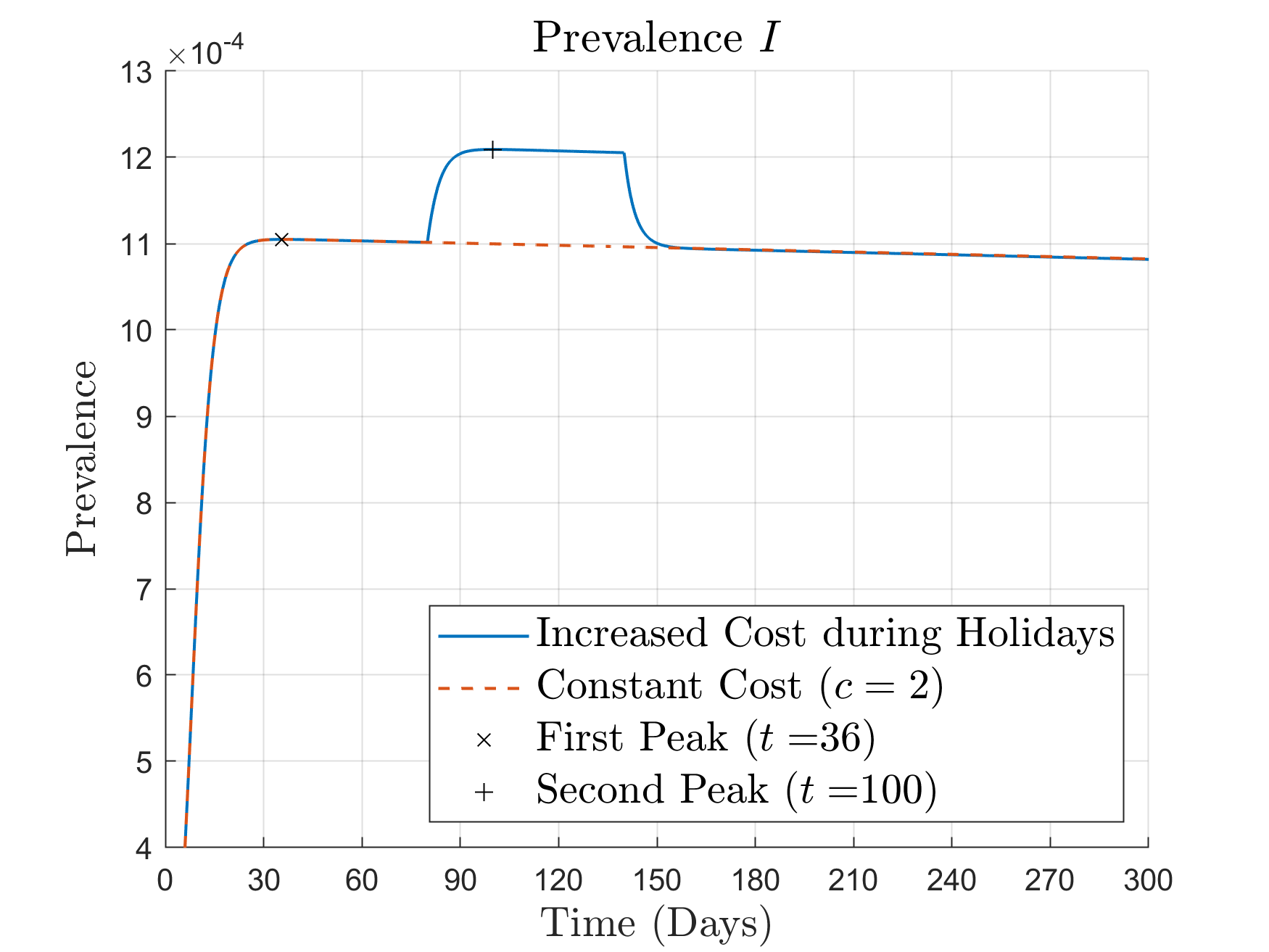}
\end{center}
\end{minipage}
\begin{minipage}{0.49 \hsize}
\begin{center}
\includegraphics[scale=0.55]{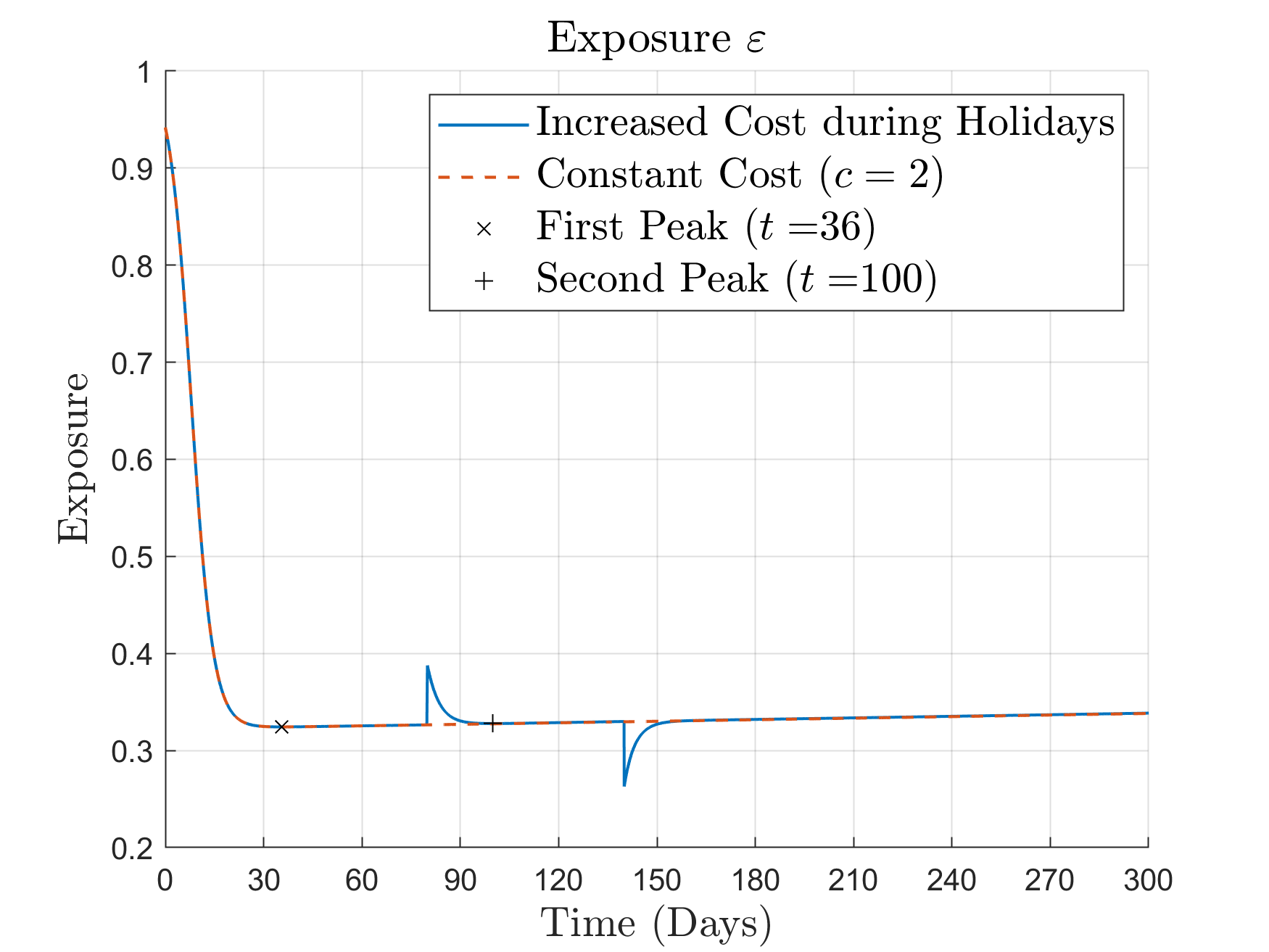}
\end{center}
\end{minipage}
\caption{\emph{Increased Distancing Cost during Holidays}. The left panel depicts the prevalence $I$ and especially the second wave. The right panel depicts the exposure level $\varepsilon$.}\label{fig:piece_wise_c_increasing} 
\end{figure}

\begin{example}
We consider the effect of holidays. We let the baseline distancing cost be $c_{0}=2$, and suppose that the distancing cost jumps to $c(t)=2.2$ from $c_{0}$ during a holiday season. Only for the purpose of illustration, the holiday season lasts for two months between days 80 and 139. One can infer from the left panel of Figure \ref{fig:const_cost_c_bar} that the new distancing cost function $c$ jumps above the original threshold $\overline{c}$ and thus induces prevalence to increase. The left panel of Figure \ref{fig:piece_wise_c_increasing} illustrates the ensuing second wave of infection. When the holiday season starts, individuals instantaneously best respond to the higher distancing cost by increasing their exposure levels. Increased exposure, in turn, leads to higher prevalence, to which then the individuals respond by decreasing their exposure again. The right panel shows that individuals' average exposure level eventually returns to the level close to (but slightly above) the case in which the distancing cost is fixed throughout.\footnote{This means that the ($10\%$) increase in the distancing cost is numerically close to the order in which the prevalence at the second wave is higher than that at the first peak.} After roughly 20 days, at around day 100, the infection peaks for the second time.\footnote{Since the slope of the prevalence curve is close to zero during the holidays, the threshold distancing cost $\overline{c}(t)$ becomes indeed close to $2.2$ during the holidays. Hence, once the holiday season finishes, the prevalence rapidly starts to decline, and in our example, towards the original prevalence level.} 
\end{example}

\subsection{Moderate Lockdown with Moderate Fatigue}\label{sec:moderate_lockdown}

\begin{example}\label{exl:c_policy_discounting}
We provide a simple illustration that the peak prevalence in the second wave may be higher than that in the first wave due to distancing fatigue. We illustrate this point for a moderate lockdown policy with moderate level of distancing fatigue. 

We let the baseline distancing cost be $c_{0}=2$, and absent a mitigation policy, the distancing cost follows equation (\ref{eq:c_distancing}) with $k=0.02$ and $r=0.05$. A social-distancing measure is introduced on day $t=30$: the distancing cost is reduced by $0.2$ (10\% of the original distancing cost).  The social-distancing measure is lifted on day $t=90$: the distancing cost is increased by $0.2$ (10\% of the original distancing cost). 

Figure \ref{fig:c_policy_discounting} depicts the threshold distancing cost $\overline{c}$, the exposure $\varepsilon$, and the prevalence $I$ with and without the distancing policy (the solid and dashed curves, respectively). The right panel shows that the second peak is higher than the first peak due to distancing fatigue. This is because, during when the social-distancing policy is implemented, individuals accumulate distancing fatigue. 

\begin{figure}
\begin{minipage}{0.32 \hsize}
\begin{center}
\includegraphics[scale=0.37]{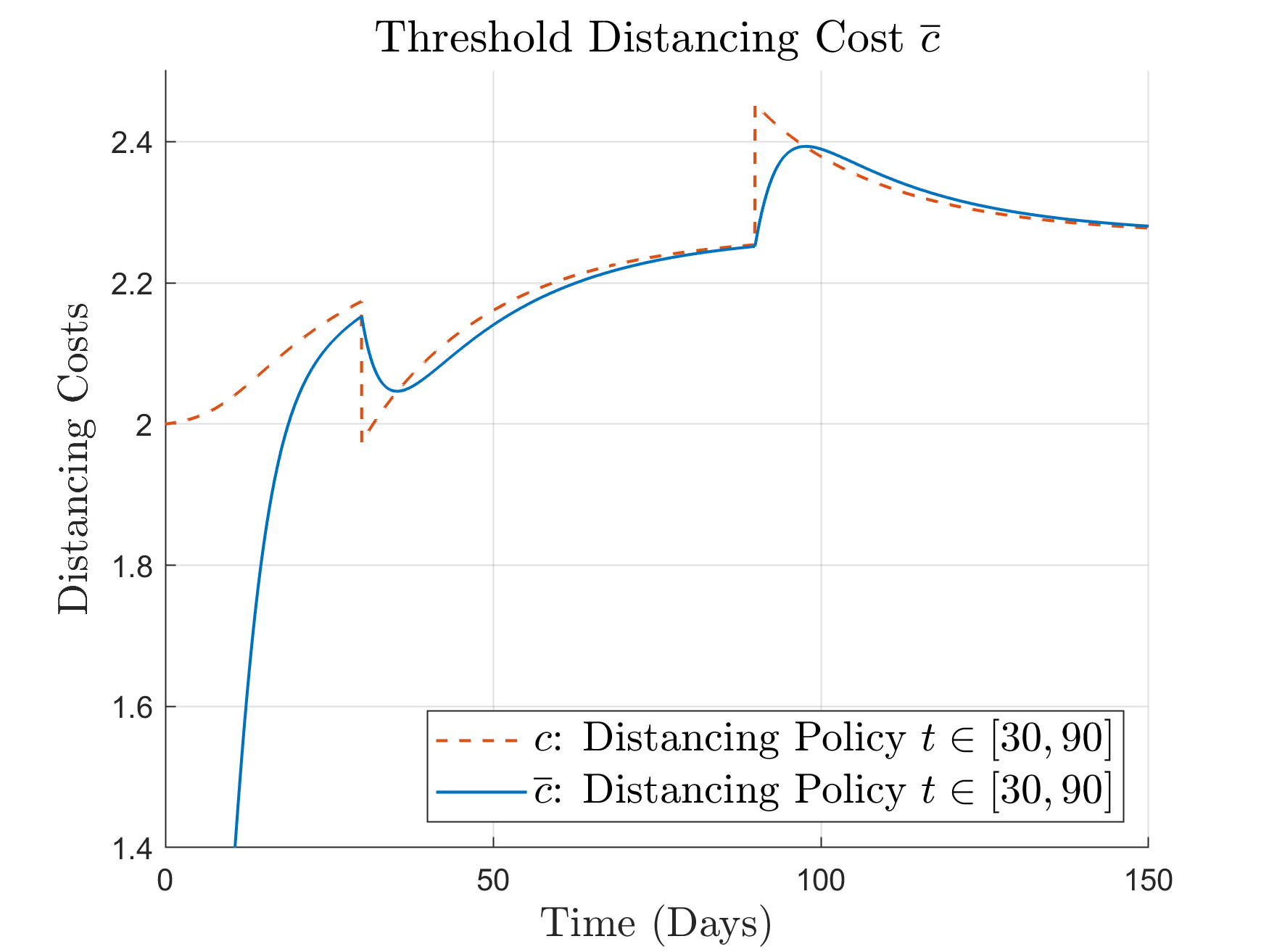}
\end{center}
\end{minipage}
\begin{minipage}{0.32 \hsize}
\begin{center}
\includegraphics[scale=0.37]{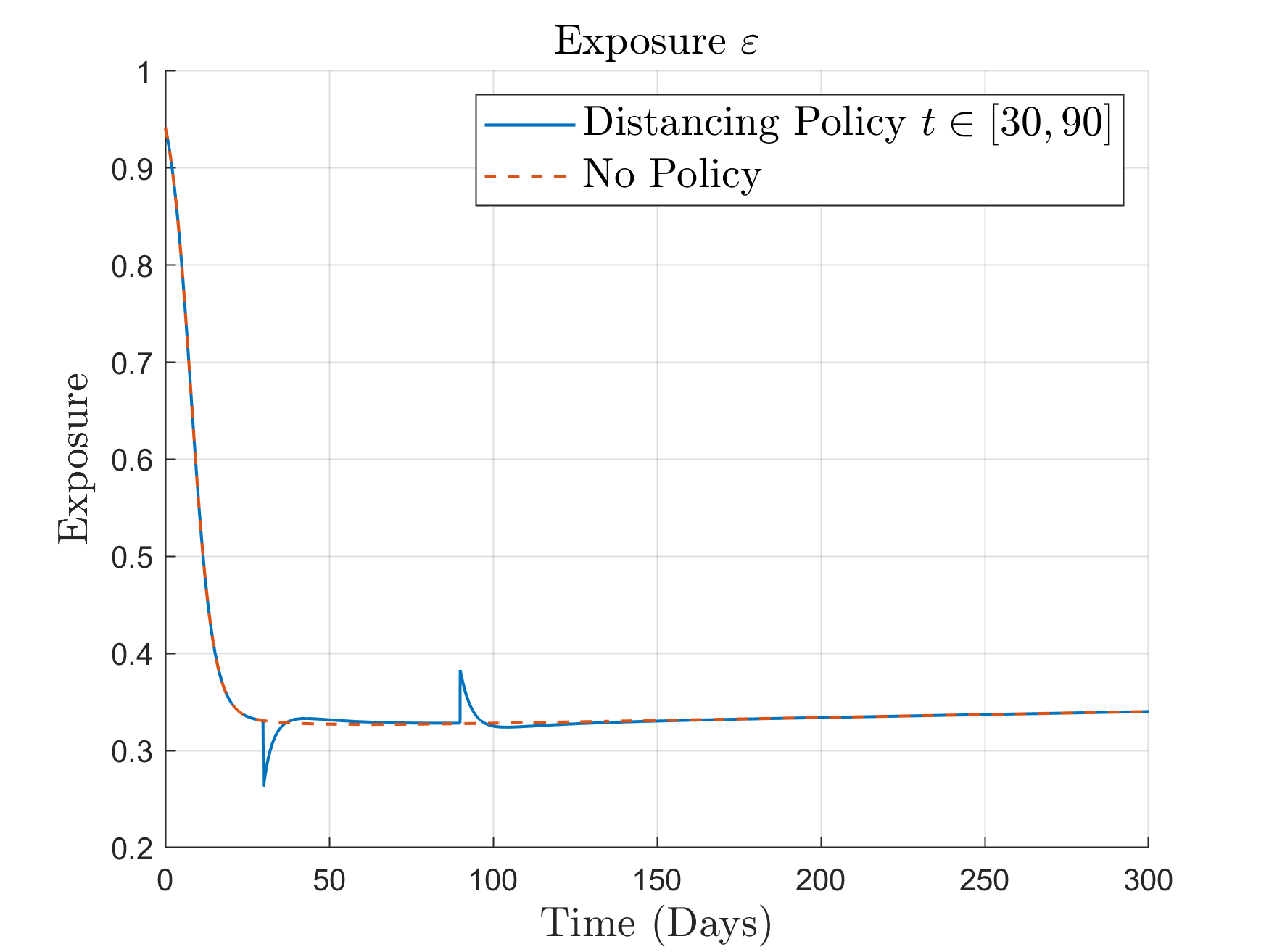}
\end{center}
\end{minipage}
\begin{minipage}{0.32 \hsize}
\begin{center}
\includegraphics[scale=0.37]{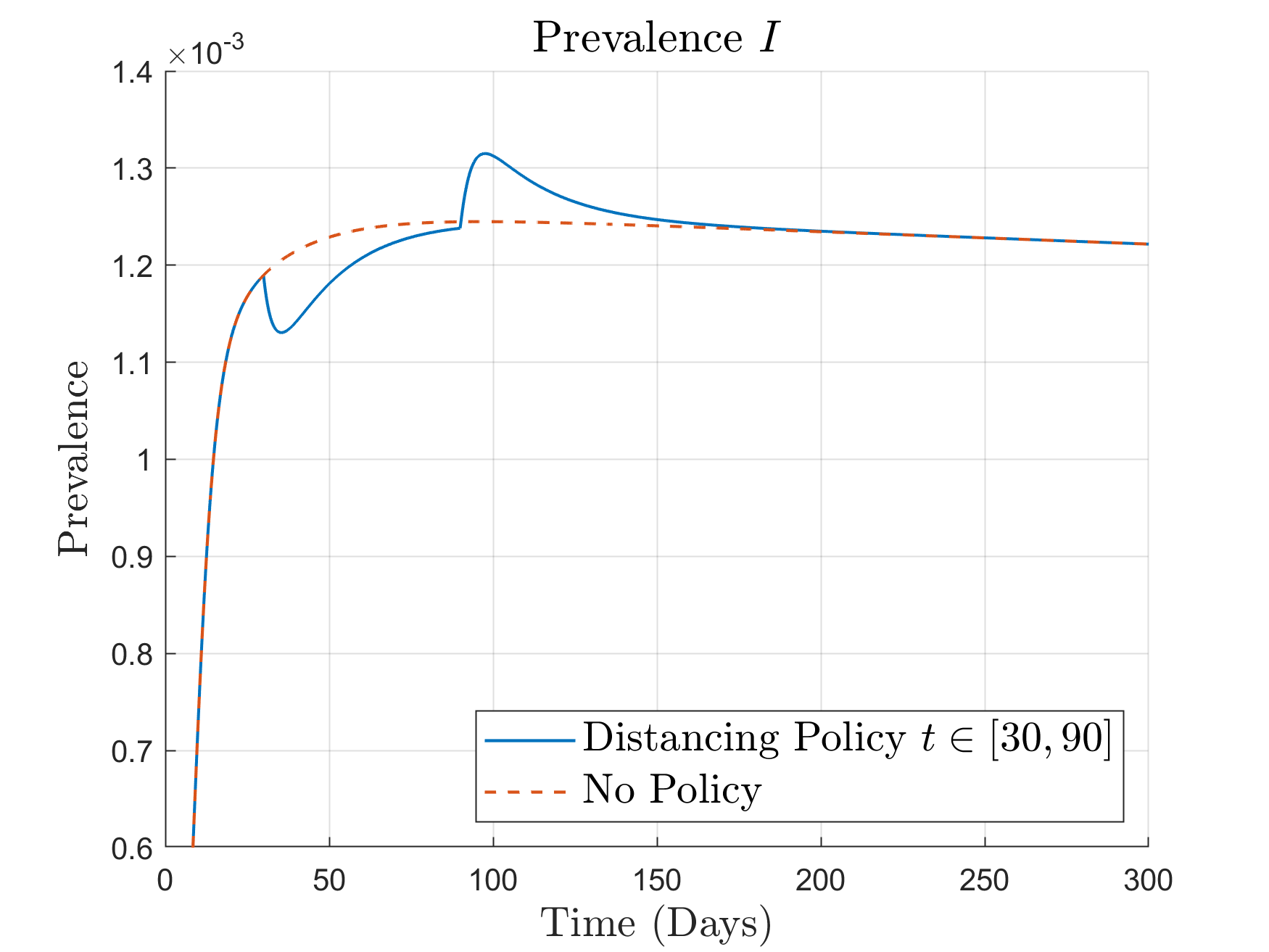}
\end{center}
\end{minipage}
\caption{\emph{Social-Distancing Policy with Distancing Fatigue}. The left panel depicts the threshold distancing cost function $\overline{c}$ over time. The central panel depicts the exposure level $\varepsilon$ over time. The right panel depicts the prevalence $I$ over time.}\label{fig:c_policy_discounting}
\end{figure}

The dashed curve (the distancing cost $c$) in the left panel of Figure \ref{fig:c_policy_discounting} shows that, soon after the implementation of the social-distancing policy by which the distancing cost is reduced, the distancing cost starts to increase. In contrast, the solid curve (the threshold distancing cost $\overline{c}$) in the left panel shows that the threshold distancing cost is endogenously decreasing soon after the implementation of the social-distancing policy. Thus, as in the right panel, the prevalence starts to increase after the implementation of the social-distancing policy due to distancing fatigue. Moreover, in the left panel, when the distancing policy is lifted on day $90$, the distancing cost $c$ is still higher than the threshold distancing cost $\overline{c}$. This suggests that, consistently with the right panel, the lifting of the distancing policy creates the higher second wave. The fact that the prevalence at the second peak is higher than the prevalence at the first peak (without the distancing policy) suggests the undesirable effect of lifting the distancing policy on the medical capacity constraint. The central panel suggests that a small increase in exposure during when the distancing policy is implemented may cause the higher second wave. 
\end{example}

\section{Far-sighted Decision-Making}\label{sec:far_sighted}

We present a model with far-sighted decision-making, and provide numerical support for our main insight: the decline of future mitigation-policy effectiveness after introducing a current policy. Hence, the assumption of myopic decision-making is not the main driver of our findings. 

As before, the individuals at each point in time decide the level of distancing, which determines the likelihood of infection. An individual's flow payoff from being in state $\theta \in \{S,I,R\}$ is $\pi_{\theta}$. We assume $\pi_{S} \geq \pi_{R} \geq \pi_{I}$.\footnote{Models with endogenous cost of infection have been presented in \citet{Reluga_10}, \citet{Fenichel_et_al_11}, \citet{Fenichel_13}, \citet*{McAdams_Song_Zou_23}, \citet{Rachel_20_Analytical}, \citet{Toxvaerd_20}, among others. Yet, analytical characterizations of equilibria even with constant distancing cost are rather elusive.} The individual discounts the future at rate $\rho >0$. 

A susceptible individual $i$ with exposure $\varepsilon_{i}(t)$ enjoys the instantaneous payoff $\pi_{S} - \frac{c_{i}(t)}{2}(1-\varepsilon_{i}(t))^{2}$. For ease of exposition, here we suppose that $\dot{c}_{i}(t)$ does not depend on the current distancing $1-\varepsilon_{i}(t)$ in (\ref{eq:distancing_cost_defi}), i.e., we suppose that the susceptible individual $i$ takes the distancing cost function $c_{i}$ as given when she decides her exposure. This is because the main insight that the effectiveness of future mitigation-policies decline after introducing a current policy holds under no distancing fatigue.\footnote{As the analysis of distancing fatigue introduces an additional state variable and thus is complicated for far-sighted individuals, we focus on the case in which each individual takes $c_{i}$ as exogenously given. Our numerical analyses confirm that our main insights carry over to far-sighted individuals. In fact, this shows an additional advantage of our myopic model when it comes to adding distancing fatigue.} 

Let $1-p_i(t)$ be the probability of being susceptible at time $t$ and, conversely, $p_i(t)$ the probability that an individual has become infected in the past. Then, $\dot{p}_i(t)$ represents the rate at which susceptible individuals become infected
\begin{equation*}
\dot{p}_i(t) = \varepsilon_i(t) \beta I(t)(1-p_i(t)),
\end{equation*}
with $p_i(0) = 0$. Since we model the behavior of susceptible individuals, the probability that they are infected at the outset is zero. Once an individual gets infected, her progression to recovery is independent of her behavior.  Her continuation payoff from the moment she became infected is $V_I= \frac{1}{\rho + \gamma} \left( \pi_{I} + \frac{\gamma}{\rho} \pi_{R} \right)$.\footnote{See \citet*[Remark 1]{Carnehl_Fukuda_Kos_23} for the formal derivation of $V_I$.}

A susceptible individual who faces average exposure $\varepsilon$ from her peers solves the problem
\begin{align}
& \max_{\varepsilon_{i}(\cdot) \in [0,1]} \int_{0}^{\infty} e^{-\rho t} \left\{ (1-p_{i}(t)) [\pi_{S}- \frac{c_{i}(t)}{2}(1-\varepsilon_{i}(t))^{2}] + p_{i}(t)\rho V_{I}  \right\} dt \label{eq:ind_problem} \\
& \text{s.t.} \notag \\
& \dot{p}_{i}(t) = \beta \varepsilon_{i}(t) I(t) (1-p_{i}(t)), \notag \\
&  p_{i}(0)=0, \notag
\end{align}
the underlying SIR dynamics given by equations (\ref{eq:S_dot}), (\ref{eq:I_dot}) and (\ref{eq:R_dot}) with the initial condition $(S(0),I(0),R(0))=(1-I_0,I_0,0)$ and $I_0\in(0,1)$, and the distancing cost function $c_{i}$ satisfying (\ref{eq:distancing_cost_defi}), where $d_{i} = 1-\varepsilon_{i}$ with $c_{i}(0)=c_{0}$.\footnote{Note that here we suppose that the individual $i$ treats $c_{i}$ as given. When $\dot{c}_{i}$ depends on $1-\varepsilon_{i}(t)$, we need to incorporate the law of motion for $c_{i}$ into the problem. Again, our assumption makes it easier to analyze the time-varying distancing costs for far-sighted individuals.} The individual's payoff can be thought of as the expected value of being susceptible or infected at each point in time where the flow payoff of an infected individual is $\rho V_I$. 

We restrict attention to distancing cost functions $c_{i}$ which satisfy 
\begin{equation}\label{eq:assumption_c}
\pi_{S} - \sup_{t \in [0, \infty)} \frac{c_{i}(t)}{2} > \rho V_{I}.
\end{equation}
This assumption states that even if a susceptible individual is fully distancing, her flow payoff of being suceptible is greater than the flow payoff of being infected.

An equilibrium $(S,I,R, c, \varepsilon, p)$ of the far-sighted decision-making model is defined analogously to our main model. Note that, in equilibrium, each $p_{i}$ is determined by $\varepsilon$, $I$, and $c$, and thus $p=p_{i}$ for each $i$. The cost of infection $\eta = \eta_{i}$, which is the co-state variable associated with the individual problem, changes over time. While $(S,I,R,c)$ is solved forward, $\eta$ is solved backward. Hence, analytically characterizing the set of equilibria is untenable. 

To characterize the forward-looking variable $\eta$, we set up the current-value Hamiltonian of problem (\ref{eq:ind_problem}): 
\begin{align}
\mathcal{H}_{i} & = (1-p_{i}(t))[\pi_{S} - \frac{c_{i}(t)}{2}(1-\varepsilon_{i}(t))^{2}] + p_{i}(t)\rho V_{I} \notag  - \eta_{i}(t)\beta \varepsilon_{i}(t) I(t)(1-p_{i}(t)), \label{eq:hamilton}
\end{align}
where $\eta_i(t)$ is the current-value co-state variable. It represents the marginal cost of an increase in the probability of being infected at time $t$. The optimality condition with respect to exposure $\varepsilon_{i}(t)$ at time $t$ is
\begin{equation*}
\frac{\partial \mathcal{H}_{i}}{\partial \varepsilon_{i}(t)} = (1-p_{i}(t)) [ c_{i}(t) (1-\varepsilon_{i}(t)) - \beta \eta_{i}(t) I(t) ] =0.
\end{equation*}
Thus, the optimality condition delivers equilibrium distancing 
\begin{equation}\label{eq:optimal_distancing}
d_i(t) =  \frac{\beta \eta_{i}(t) I(t)}{c_{i}(t)},
\end{equation}
provided that the entire distancing path admits an interior solution, i.e., that $d_i(t)\in[0,1]$ for all $t$. One should keep in mind that the marginal cost of an increased probability of infection, $\eta_i(t)$, is positive due to the assumption given by (\ref{eq:assumption_c}). The current-value co-state variable $\eta_{i}$ follows the adjoint equation
\begin{align*}
\dot{\eta}_{i}(t) & = \rho \eta_{i}(t) + \frac{\partial \mathcal{H}_{i}}{\partial p_{i}(t)} \notag \\
& = \eta_{i}(t) \left( \rho + \varepsilon_{i}(t) \beta I(t) \right) + \left( \pi_{S} - \frac{c_{i}(t)}{2}(1-\varepsilon_{i}(t))^{2}  - \rho V_{I}\right). 
\end{align*}
The transversality condition is $\displaystyle \lim_{t \rightarrow \infty} e^{-\rho t}\eta_{i}(t) = 0$. In equilibrium, $\eta = \eta_{i}$ for all $i$. 

Using the adjoint equation and the transversality condition, we can solve for $\eta$. 

\begin{lmm}\label{lemma:eta}
Suppose that the rest of the population is following the strategy $\varepsilon$, and $\varepsilon_{i}$ is the individual $i$'s best response. Then
\begin{equation*} 
\eta_i(t) = \int_{t}^{\infty} e^{-\rho (s-t)} \frac{1-p_i(s)}{1-p_i(t)} \left( \pi_{S} -  \frac{c_{i}(t)}{2}(1-\varepsilon_i(s))^2 - \rho V_{I} \right) ds.
\end{equation*}
Let $(S, I, R, \varepsilon, p)$ be an equilibrium. Then
\begin{equation*}
\eta(t) = \int_{t}^{\infty} e^{-\rho (s-t)} \frac{S(s)}{S(t)} \left( \pi_{S} -  \frac{c(s)}{2}(1-\varepsilon(s))^2 - \rho V_{I} \right) ds.
\end{equation*} 
\end{lmm}

The proof of this lemma is similar to that of \citet*[Lemma 2]{Carnehl_Fukuda_Kos_23}, and thus it is omitted. Instead, we provide the interpretation of the lemma. We term $\pi_{S} - \frac{c(t)}{2}(1-\varepsilon(t))^{2} - \rho V_{I}$ the \emph{susceptibility premium} at time $t$. It is the difference in flow payoffs between being susceptible and being infected. The cost of getting infected, $\eta(t)$, is the discounted value of the susceptibility premium over time weighted by the conditional probability of being susceptible at each time in the future, $s\geq t$, $\frac{S(s)}{S(t)}$. Distancing over a period of time reduces the quality of life and, thus, the susceptibility premium. However, it also decreases the probability that the individual will get infected and rewards her with the premium for a longer period of time.

As in \citet*[Lemma 3]{Carnehl_Fukuda_Kos_23}, one can also show that $\eta$ is bounded. Letting $(S, I, R, c, \varepsilon, p)$ be an equilibrium, 
\begin{equation*}
\displaystyle \frac{\pi_S - \rho V_I-\frac{c(t)}{2}}{\rho+\beta} \leq \eta (t) \leq \frac{\pi_S-\rho V_{I}}{\rho} \text{ and } \lim_{t \rightarrow \infty} \eta(t) = \frac{\pi_{S} - \rho V_{I}}{\rho}. 
\end{equation*}
As time passes, $\eta$ eventually converges to the upper bound, which is attained when individuals choose full exposure in perpetuity without facing any risk of becoming infected. This is the case in which getting infected would be most costly as there is no need to distance and no risk of future infection. The convergence to this bound is intuitive, because the disease dies out and obviates the need for distancing in the limit as time goes to infinity. 

Below, we present the results of our numerical simulations with endogenous $\eta$. We calibrate the parameters for the beginning of COVID-19 (recall footnote \ref{fn:calibration} in the main text). The value of $\eta$ in the main text corresponds to the upper bound $\eta = \frac{\pi_{S} - \rho V_{I}}{\rho}$.\footnote{Hence, by assumption, in our numerical simulations of the model in which $\eta$ is fixed at the upper bound, individuals engage distancing more and the prevalence is lower. In contrast, by taking the lower bound of $\eta$, we can also bound the prevalence from above. This way, the model with constant infection cost can also shed light on the dynamics of the model with endogenous infection cost.} For the values of $\pi_{S}$, $\rho$, and $V_{I}$, see \citet*{Carnehl_Fukuda_Kos_23}.

\begin{figure}[t]
\begin{minipage}{0.32 \hsize}
\begin{center}
\includegraphics[scale=0.38]{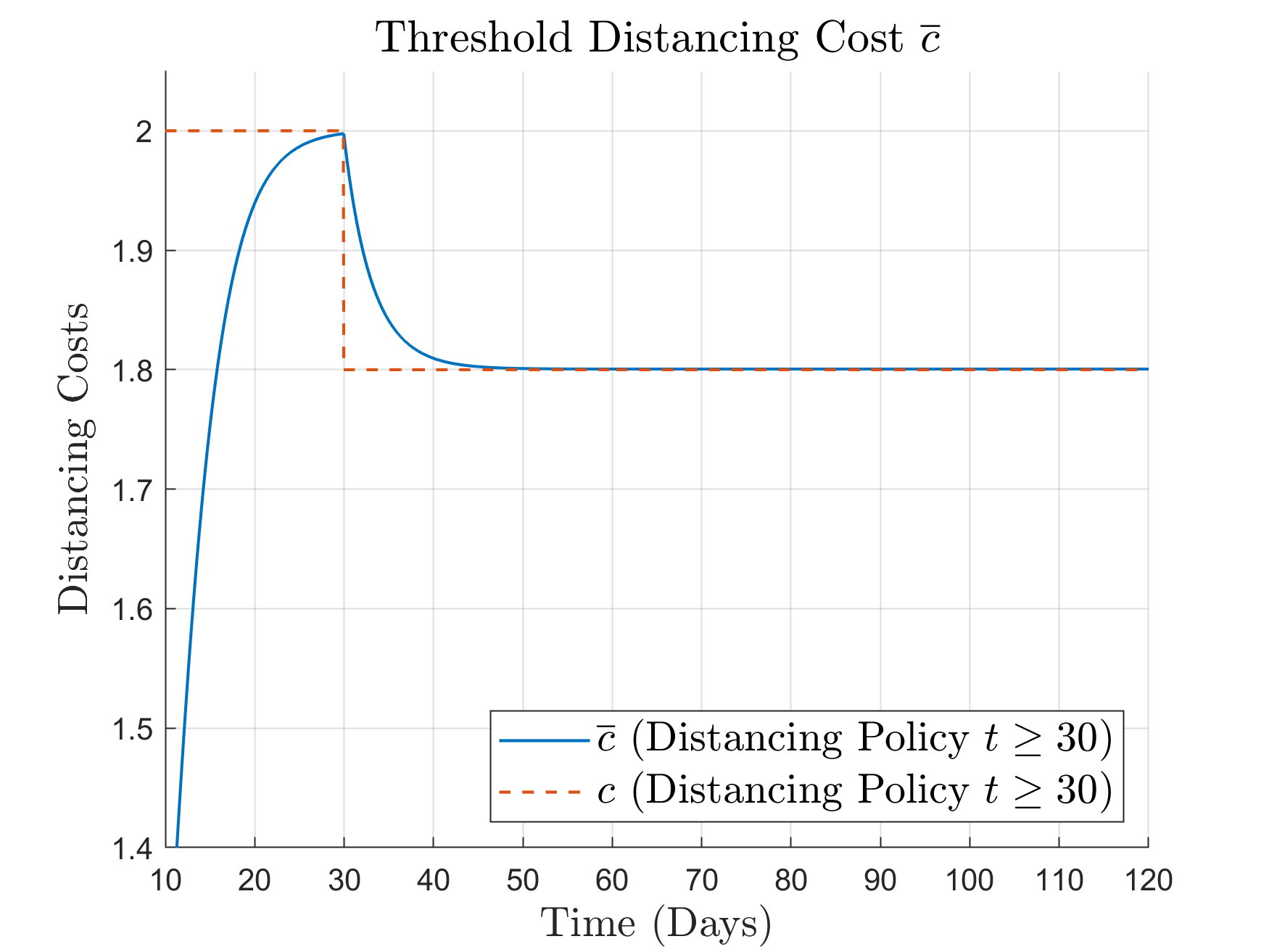}
\end{center}
\end{minipage}
\begin{minipage}{0.32 \hsize}
\begin{center}
\includegraphics[scale=0.38]{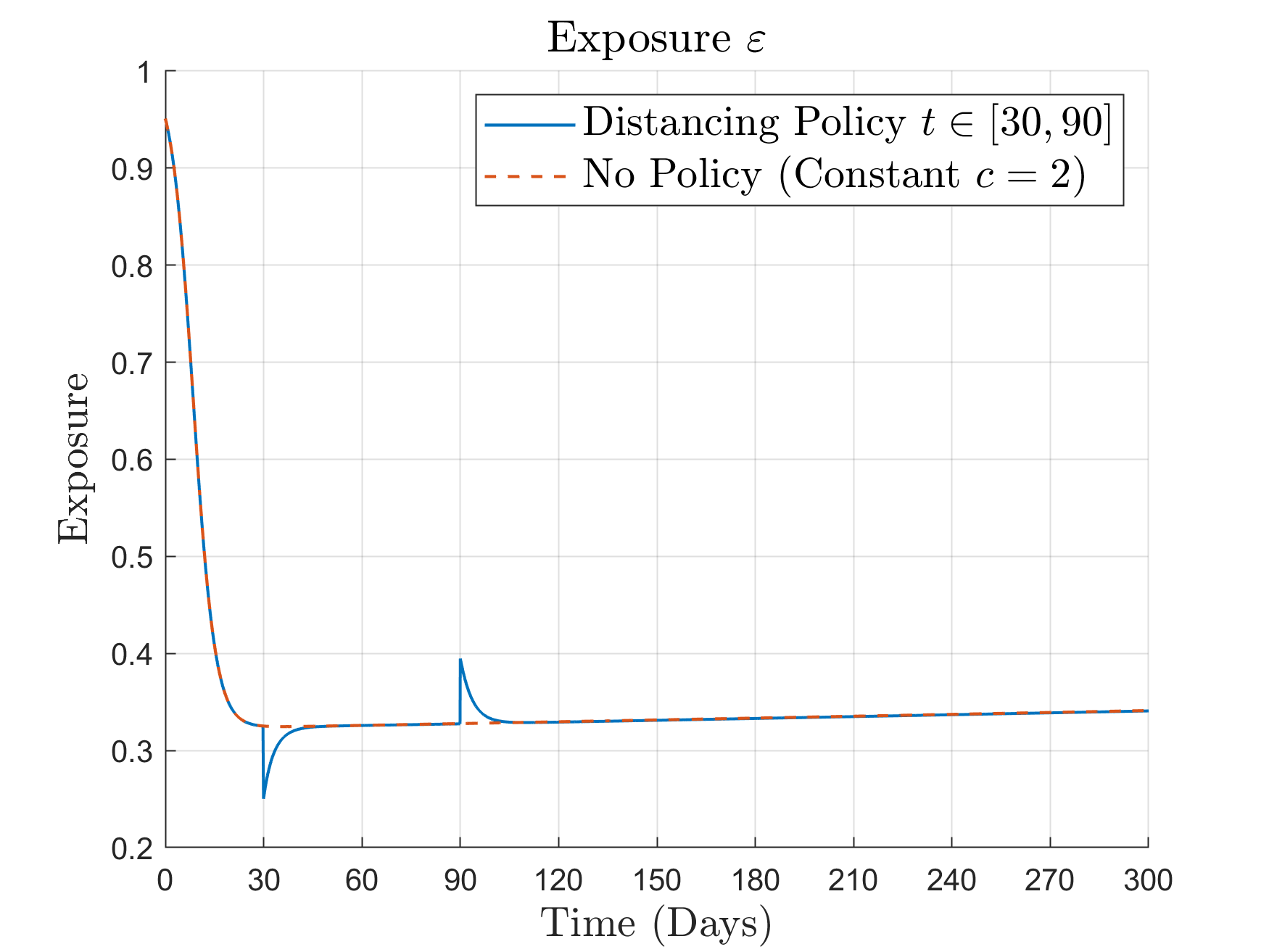}
\end{center}
\end{minipage}
\begin{minipage}{0.32 \hsize}
\begin{center}
\includegraphics[scale=0.38]{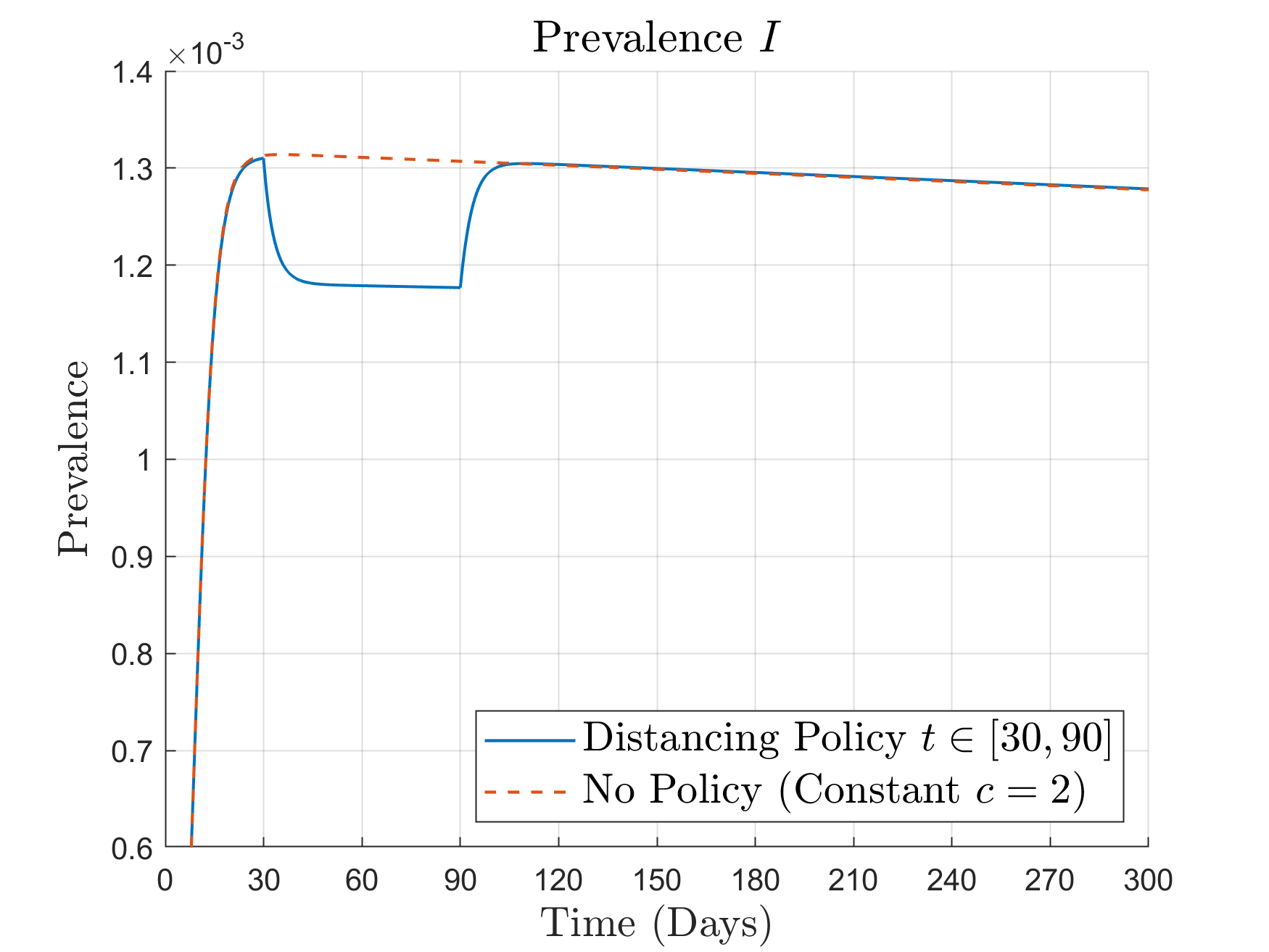}
\end{center}
\end{minipage}
\caption{\emph{Social-Distancing Policy}. The left panel depicts the threshold distancing cost function $\overline{c}$ over time. The central panel depicts the exposure level $\varepsilon$ over time. The right panel depicts the prevalence $I$ over time.}\label{fig:piece_wise_c_decreasing_far_sighted}
\end{figure}

Specifically, we revisit Example \ref{exl:temporary_lockdown} with endogenous $\eta$. To that end, we define 
\begin{equation*}
\overline{c}(t) := \begin{cases} 
\frac{\beta^2 I(t) S(t) \eta(t)}{\beta S(t)-\gamma}, & \text{if } S(t)>\frac{\gamma}{\beta}  \\ \infty, & \text{if } S(t) \leq \frac{\gamma}{\beta} 
\end{cases}.
\end{equation*}
Note that the only difference from the main text is that $\eta$ is now time-varying. This is because the exposure level of the far-sighted individual is $\varepsilon(t) = 1 - \frac{\beta \eta(t)I(t)}{c(t)}$. One can show that Proposition \ref{prop:threshold_cost} holds under this setting, as the proof in Appendix \ref{sec:proofs} simply extends to this case. 

The left panel of Figure \ref{fig:piece_wise_c_decreasing_far_sighted} illustrates the threshold distancing function $\overline{c}$. Figure \ref{fig:piece_wise_c_decreasing_far_sighted} look similar to Figure \ref{fig:piece_wise_c_decreasing}. 

To sum up, while one might argue that individuals do not fully discount the future, the difficulty in predicting the path of an epidemic might induce them to simply respond to the current state of the epidemic. In addition, analytical results for the SIR model with endogenous distancing by far-sighted individuals are few and far between.\footnote{In the context of behavioral SIR models, it has not even been established whether such a model with equilibrium distancing has a single peak even for the case in which distancing cost is constant over time. For example, in an optimal planner problem of distancing, \citet{Kruse_Strack_22} show that the prevalence peaks at most twice.} The myopic model enables one to provide analytical insights and pave the road towards the understanding of the model with far-sighted individuals. Moreover, our numerical analysis highlights that assuming a fixed cost of infection is not the main driver of our findings. 

\end{document}